\documentclass[aps,prb,twocolumn,showpacs,preprintnumbers,amsmath,amssymb,superscriptaddress]{revtex4}
\usepackage{dcolumn}
\usepackage{bm}
\usepackage{graphicx}
\usepackage{times}
\usepackage{epstopdf}
\usepackage{color}
\usepackage[subrefformat=parens,labelformat=parens]{subfig}

\makeatletter
\newcommand*{\citenumns}[2][]{%
  \begingroup
  \let\NAT@mbox=\mbox
  \let\@cite\NAT@citenum
  \let\NAT@space\NAT@spacechar
  \let\NAT@super@kern\relax
  \renewcommand\NAT@open{}%
  \renewcommand\NAT@close{}%
  \cite[#1]{#2}%
  \endgroup
}
\makeatother
\begin{document}

\definecolor{Red}{rgb}{1,0,0}
\definecolor{Blu}{rgb}{0,0,01}
\definecolor{Green}{rgb}{0,1,0}
\newcommand{\red}{\color{Red}}
\newcommand{\blu}{\color{Blu}}
\newcommand{\green}{\color{Green}}
\newcommand{\rvec}{\mathbf{r}}
\newcommand{\nvec}{\mathbf{n}}
\newcommand{\ev}[1]{\langle#1\rangle}
\newcommand{\abs}[1]{\left|#1\right|}

\title{Fluctuation effects in rotating Bose-Einstein condensates with broken $\mathrm{SU}(2)$ and
$\mathrm{U}(1)\times \mathrm{U}(1)$ symmetries in the presence of intercomponent density-density
interactions.}
\author{Peder Notto Galteland}
\affiliation{Department of Physics, Norwegian University of Science and Technology, N-7491
Trondheim, Norway}
\author{Egor Babaev}
\affiliation{Department of Theoretical Physics, The Royal Institute of Technology, 10691 Stockholm, Sweden}
\author{Asle Sudb\o}
\affiliation{Department of Physics, Norwegian University of Science and Technology, N-7491
Trondheim, Norway}
\date{\today}

\begin{abstract} Thermal fluctuations and melting transitions for rotating single-component
  superfluids have been intensively studied and are well understood. In contrast, the thermal
  effects on vortex states for two-component superfluids with density-density interaction, which
  have a much richer variety of vortex ground states, have been much less studied. Here, we
  investigate the thermal effects on vortex matter in superfluids with $\mathrm{U(1)}\times
  \mathrm{U(1)}$ broken symmetries and intercomponent density-density interactions, {as well as the
    case with a larger $\mathrm{SU(2)}$ broken symmetry obtainable from the $\mathrm{U(1)}\times
  \mathrm{U(1)}$-symmetric case by tuning scattering lengths}. In the former case we find that, in
  addition to  first-order melting transitions, the system exhibits thermally driven phase
  transitions between square and hexagonal lattices. Our main result, however, concerns the case
  where the condensate exhibits $\mathrm{SU(2)}$-symmetry, and where vortices are not topological.
  At finite temperature, the system exhibits effects which do not have a counter-part in single
  component systems. Namely, it has a state where thermally averaged quantities show no regular
  vortex lattice, yet the system retains superfluid coherence along the axis of rotation. In such a
  state, the thermal fluctuations result in transitions between different (nearly)-degenerate vortex
  states without undergoing a melting transition. Our results apply to multi-component Bose-Einstein
  condensates, and we suggest how to experimentally detect some of these unusual effects in such
  systems.

\end{abstract}
\maketitle

\section{Introduction}
\label{sec:intro}

Bose-Einstein condensates (BECs) with a multicomponent order parameter, and the topological defects
such systems support, represent a topic of great current interest in condensed matter
physics.\cite{Wieman97, Cornell98, Cornell99, Modugno2002, mueller02, Kasamatsu2003, Cornell2004,
Papp2008, Thalhammer2008, makoto2, Tojo2010, McCarron2011, mottonen, tsubota3, cipriani} Such
multicomponent condensates may be realized as mixtures of different atoms, mixtures of different
isotopes of an atom, or mixtures of different hyperfine spin states of an atom. The interest in such
condensates from a fundamental physics point of view is mainly attributed to the fact that one may
tune various interaction parameters over a wide range in a BEC.
This enables the study of a variety of physical effects which are not
easily observed in other superfluid systems such as He$^3$ and He$^4$.

The behavior of a single-component BEC under rotation is well
known. The ground state is a hexagonal lattice of vortex defects which melts to
a vortex liquid via a first-order phase transition.  This is well described
by the London model, where amplitude fluctuations may be
ignored. Over the years, in the context of studying vortex lattice melting in high-$T_c$ superconductors,
many works have confirmed this through numerical Monte Carlo simulations for systems in the frozen gauge,
three-dimensional (3D) $XY$ and Villain approximations, \cite{Teitel91,Sudbo92,Teitel97,Hu97,Nguyen98_1,Nguyen98_2,Stroud98,Nguyen99,Chin99,Nguyen_EPL99} as
well as in the lowest-Landau-level approximation,\cite{Nordborg97} and by mapping it to a model of
2D bosons.\cite{MacDonald97} Single component condensates have been available experimentally
  for quite some time\cite{Cornell95,Ketterle95}, and the hexagonal lattice ground state has been
verified.\cite{Abo-Shaeer2001}

Condensates with two components of the order parameter have also been studied extensively.
Analytical works focusing on determining the $T=0$ ground states have demonstrated interesting vortex solutions and  a range of
unusual lattice structures\cite{mueller02,
Kasamatsu2003,makoto2,tsubota3,koba,mottonen,catelani}. By varying the ratio between inter- and
intra-component couplings, the ground-state lattice undergoes a structural change from hexagonal
symmetry through square symmetry to double-core lattices and interwoven sheets of vortices. Similar
systems with three components have also been studied.\cite{cipriani} Experimentally, spinor condensates
have been realized in two general classes of systems. The first option is to use one species of atoms,
usually rubidium, and prepare it in two separate hyperfine spin states.\cite{Wieman97,Cornell98}
Vortices\cite{Cornell99} and vortex lattices\cite{Cornell2004} have been realized in these binary mixtures,
where both hexagonal and square vortex lattice states were observed. The other option is to mix condensates of two different
species of atoms.\cite{Modugno2002,McCarron2011} The use of Feshbach resonances\cite{Feshbach,Inouye1998}
allows direct tuning of the scattering lengths, and by extension the inter- and intracomponent
interactions of multicomponent condensates.\cite{Papp2008,Thalhammer2008, Tojo2010}

In this paper, we consider a specific model of a two-component BEC, which has the full range of
fluctuations of the order parameter field included, as well as intercomponent density-density
interactions. We consider the model with $\mathrm{U(1)}\times\mathrm{U(1)}$- and as
$\mathrm{SU(2)}$ symmetries. For the $\mathrm{U}(1) \times \mathrm{U}(1)$ case, we find a succession
of square and hexagonal vortex ground-state patterns as the intercomponent interaction strength is
varied, along with the possibility of thermal reconstruction from a square to a hexagonal vortex
lattice as temperature is reduced.

The $\mathrm{SU}(2)$-symmetric case is interesting and experimentally realizable. In this case
$\mathrm{U}(1)$-vortices are no longer topological, in contrast to the $\mathrm{U(}1) \times
\mathrm{U}(1)$-symmetric case. In this case, when fluctuation effects are included we find a highly
unusual vortex state where there is no sign of any vortex lattice.  Nonetheless, global phase
coherence persists. This state of vortex matter is a direct consequence of massless
amplitude fluctuations in the order parameter, when the broken symmetry of the system is
$\mathrm{SU}(2)$. At the $\mathrm{SU}(2)$ point, but at lower temperatures, we also observe
dimerized vortex ground-state patterns.

The paper is organized as follows. The model and definitions of relevant quantities are presented
in Section~\ref{sec:model}. The technical details of the Monte Carlo simulations are
briefly considered in Section~\ref{sec:MCdetails}. In Section~\ref{sec:results}, the results are
presented and discussed.  In Section~\ref{sec:exp}, we discuss how to experimentally verify the results
we find. Some technical details, and the investigation of the order of the melting transitions
with full amplitude distributions included, for the cases $N=1$ and $N=2$, are relegated to Appendixes.

\section{Model and definitions}
\label{sec:model}

In this section we present the model used in the paper, first in a continuum description and then on a
three-dimensional cubic lattice appropriate for Monte Carlo simulations. The relevant quantities for
the discussion are also defined.

\subsection{Continuum model}
We consider a general Ginzburg-Landau(GL) model of an $N$-component Bose-Einstein condensate, coupled to a uniform
external field, which in the thermodynamical limit is defined as

\begin{equation} \mathcal{Z}=\int\prod_i^N\mathcal{D}\psi_i^\prime
\text{e}^{-\beta H},
\end{equation}
where
\begin{align}
  H=\int
  d^3r\Bigg[&\sum_{i=1}^N\sum_{\mu=1}^3\frac{\hbar^2}{2m_i}\left|(\partial_\mu-\mathrm{i}\frac{2\pi}{\Phi_0}A_\mu^\prime)\psi_i^\prime\right|^2\nonumber\\
                            &+\sum_i^N\alpha_i^\prime\abs{\psi_i^\prime}^2+
\sum_{i,j=1}^Ng_{ij}^\prime\abs{\psi_i^\prime}^2\abs{\psi_j^\prime}^2\Bigg]
  \label{eq:genH}
\end{align}
is the Hamiltonian. Here, the field $A_\mu^\prime$ formally appears as a non-fluctuating gauge-field
and parametrizes the angular velocity of the system. The fields $\psi_i^\prime$ are dimensionful
complex fields, $i$ and $j$ are indices running from $1$ to $N$ denoting the component of the order
parameter (a ``color" index), $\alpha_i^\prime$ and $g_{ij}^\prime$ are Ginzburg-Landau parameters,
$\Phi_0=h/2e$ is the the coupling constant to the rotation induced vector potential, and $m_i$ is
the particle mass of species $i$. For mixtures consisting of different atoms or different isotopes
of one atom, the masses will depend on the index $i$, while for mixtures consisting of atoms in
different hyperfine spin states, the masses are independent of $i$.  The inter- and intracomponent
coupling parameters $g_{ij}^\prime$ are related to real inter- and intracomponent scattering
lengths, $a_{ij}$, in the following way
\begin{align}
  g_{ii}^\prime &= \frac{4\pi\hbar^2 a_{ii}}{m_i},\\
  g_{ij}^\prime &= \frac{8\pi\hbar^2 a_{ij}}{m_{ij}}, (i\neq j)
\end{align}
where $m_{ij}=m_i~m_j/(m_i+m_j)$ is the reduced mass. In this paper we focus on using BECs of
homonuclear gases with several components in different hyperfine states, hence $m_i=m\,\forall\, i$.
Inter-component drag in BEC mixtures has been considered in previous works using Monte-Carlo simulation
(ignoring amplitude fluctuations), but we will not consider this case here.~\cite{Dahl2008_2,Dahl2008,Dahl2008_3,2004PhRvL..92c0403K,2004PhRvL..93w0402K}

We find it convenient for our purposes to rewrite~\eqref{eq:genH} on the following form, the details of which are relegated to
Appendix~\ref{app:rewrite},

\begin{equation}
  H = \int d^3r\Bigg[\frac{1}{2}(D_\mu\Psi)^\dag(D_\mu\Psi)+V(\Psi)\Bigg].
  \label{eq:Ham}
\end{equation}
Here, $\Psi$ is an $N$-component spinor of dimensionless complex fields, which consists of an
amplitude and a phase, $\psi_i=\abs{\psi_i}\exp{(i\theta_i)}$, $D_\mu =
\partial_\mu-\mathrm{i}\frac{2\pi}{\Phi_0}A^\prime_\mu$ is the covariant derivative, and summation over repeated spatial
indices is implied.  We neglect, for simplicity, the presence of a trap and the centrifugal part of the
potential. We consider only the case where the vector potential is
applied to each component of $\Psi$, as follows from the fact that the masses are independent of
species-index $i$.

We have studied this model in detail with $N=2$, where we write the potential in the form
\begin{equation}
V(\Psi)=\eta(\left|\Psi\right|^2-1)^2+\omega(\Psi^\dag\sigma_z\Psi)^2.
\label{eq:N2pot}
\end{equation}
This formulation is more relevant for our discussion, as it immediately highlights the symmetry of
$\Psi$, as well as the soft constraints applied to it. The details of the reparametrization are shown
in Appendix~\ref{app:rewrite}.

Note that Eq. \ref{eq:N2pot} may also be rewritten in the form (correct up to an additive constant term)
\begin{equation}
V=(\eta+\omega)(\left|\psi_1\right|^4+\left|\psi_2\right|^4)+2(\eta-\omega)\left|\psi_1\right|^2\left|\psi_2\right|^2
\label{eq:inter_intra_g}.
\end{equation}
Comparing with Eq. \ref{eq:genH}, we have $g_{11}=g_{22}\equiv g=\eta+\omega$ and $g_{12}=\eta-\omega$. The model features
repulsive inter-component interactions provided $ \eta - \omega > 0$, and this is the case we will mainly focus on. We will
however briefly touch upon the case  $\eta - \omega < 0$ corresponding to an attractive  inter-component density-density
interaction, which leads to ground states with overlapping vortices in components $1$ and $2$. Normalizability of the individual
order parameter components, or equivalently boundedness from below of the free energy, requires that $\eta + \omega> 0$. Thus,
while $\omega > \eta$ makes physical sense, $\omega < -\eta$ does not. In this paper, we assume
$\eta > 0$ and {$\omega \geq 0$.}

Two-component BECs feature considerably richer physics than a single-component BEC. Since the gauge-field
parametrizing the rotation of the system is non-fluctuating, there is no gauge-field-induced current-current interaction
between the two condensates (unlike in multi-component superconductors). The only manner in which the two superfluid
condensates interact is via the inter-component density-density interaction  $2 (\eta-\omega) |\psi_1|^2 |\psi_2|^2$. In
the limit where the amplitudes of each individual component are completely frozen and uniform throughout the
system, one recovers the physics of two decoupled 3D$XY$ models, with a global $\mathrm{U(1)}\times
\mathrm{U(1)}$ symmetry. The density-density interaction between $\psi_1$ and $\psi_2$ leads to
interactions between the topological defects excited in each component. As a result, a first
order melting of two decoupled hexagonal lattices is not the only possible phenomenon that could
take place. Previous experiments and numerical studies have reported a structural change of the ground state from a hexagonal
to a square lattice of vortices as the effective inter-component coupling is increased.\cite{mueller02,Cornell2004,Kasamatsu2003}
This corresponds to increasing the ratio $\eta/\omega$ in our case. As we will see below, other
unusual phenomena can also occur, notably when thermal fluctuations are included.

One special case of the model deserves some extra attention. If one takes the limit $\omega
\rightarrow 0$ in Eq.~\eqref{eq:N2pot} the symmetry of the model is expanded to a global
$\mathrm{SU(2)}$ symmetry. One may then shift densities from one component to the other with
impunity, as long as $|\psi_1|^2+|\psi_2|^2$ is left unchanged. This effectively leads to massless
amplitude-fluctuations in the components of the order parameter. Therefore, it is possible to unwind
a $2 \pi$ phase winding in one component by letting the amplitude of the same component vanish.
The introduction of this higher symmetry leads to very different vortex ground states than what are
found in the $\mathrm{U(1)}\times \mathrm{U(1)}$-symmetric case with $\omega \neq 0$.

\subsection{Separation of variables}

In multicomponent GL models complex objects, such as combinations of vortices
of different colors, are often of interest. In general, it is possible to
rewrite an $N$-component model coupled to a gauge field, fluctuating or not, in
terms of one mode coupled to the field and $N-1$ neutral
modes.\cite{Smiseth2005,Herland2010} For a more general discussion of charged
and neutral modes in the presence of amplitude fluctuations see Refs.
\citenumns{2002PhRvB..65j0512B} and \citenumns{CPN}. Considering only the
kinetic part of the two-component Hamiltionian, $H_k$, we have the following
expression:

\begin{align}
  H_k=&\frac{1}{2\left|\Psi\right|^2}\left|\psi_1^*\partial_\mu\psi_1+\psi_2^*\partial_\mu\psi_2-\mathrm{i}A_\mu\left|\Psi\right|^2\right|^2\nonumber\\
+&\frac{1}{2\left|\Psi\right|^2}\left|\psi_1\partial_\mu\psi_2-\psi_2\partial_\mu\psi_1\right|^2.
\end{align}
Hence, the first mode couples to the applied rotation, while the second does not. This corresponds
to the phase combinations $\theta_1+\theta_2$ and $\theta_1-\theta_2$, respectively.

\subsection{Lattice regularization}

In order to perform simulations of the continuum model, we define the field $\Psi$ on a discrete set of
coordinates, \textit{i.e} $\Psi(\mathbf{r})\rightarrow\Psi_\mathbf{r}$, where
$\rvec\in(i\hat{\mathbf{x}}+j\hat{\mathbf{y}}+k\hat{\mathbf{z}} | i,j,k=1,\ldots,L)$. Here, $L$ is
the linear size in all dimensions; the system size is $V=L^3$. We use periodic boundary conditions
in all directions. By replacing the differential operator by a gauge-invariant forward difference

\begin{equation}
\bigg(\frac{\partial}{\partial
r_\mu}-\mathrm{i}A_\mu(\mathbf{r})\bigg)\Psi(\mathbf{r})\rightarrow\frac{1}{a}\bigg(\Psi_{\mathbf{r}+a\hat{\boldsymbol{\mu}}}\text{e}^{-\mathrm{i}\frac{2\pi}{\Phi_0}aA^\prime_{\mu,\mathbf{r}}}-\Psi_\mathbf{r}\bigg),
\end{equation}
and introducing real phases and amplitudes
$\psi_{\rvec,i}=\left|\psi_{\rvec,i}\right|e^{i\theta_{\rvec,i}}$ we can rewrite the
Hamiltonian.

\begin{align}
  H=&\sum_{\substack{\rvec,\hat{\boldsymbol{\mu}}\\i}}\big|\psi_{\rvec+\hat{\boldsymbol{\mu}},i}\big|\big|\psi_{\rvec,i}\big|
                                   \cos(\theta_{\rvec+\hat{\boldsymbol{\mu}},i}-\theta_{\rvec,i}-A_{\mu,\rvec})\nonumber\\
  +&\sum_\rvec V(\Psi_\rvec).
\end{align}

The lattice spacing is chosen so that it is smaller than the relevant length scale of
variations of the amplitudes.
A dimensionless vector potential, $A_\mu$, has also been introduced.
See Appendix~\ref{app:rewrite} for details.  We denote
the argument of the cosine as $\chi_{\rvec,i}^\mu$, as a shorthand.

\subsection{Observables}
An important and accessible quantity when exploring phase transitions is the specific heat of the system,
\begin{equation}
    c_V=\beta^2\frac{\langle H^2\rangle-\langle H\rangle^2}{L^3}.
\end{equation}
While crossing a first order transition there is some amount of latent heat in the system, manifesting
itself as a $\delta$-function peak of the specific heat in the thermodynamic limit. On the lattice one
expects to see a sharp peak, or anomaly, at the transition. This is used to characterize the
transition as first order.

A useful measure of the global phase coherence of the system, is the helicity modulus, which is
proportional to the superfluid density. It serves as a probe of the transition from a superfluid to
a normal fluid. In the disordered phase, the moduli in all directions are zero, characterizing an
isotropic normal-fluid phase. The cause of this is a vortex loop blowout. Moving to the ordered
phase, all moduli evolve to a finite value. If we turn on the external field we still have zero
coherence in all directions in the disordered phase. In the ordered phase, however, the helicity
modulus along the direction of the applied rotation jumps to the finite value through a first order
transition. The value of the transverse moduli will remain zero. Formally, the helicity modulus is
defined as a derivative of the free energy with respect to a general, infinitesimal phase twist
along $r_\mu$,\cite{Fisher73}. That is, we perform the replacement

\begin{equation}
  \theta_{\rvec,i}\rightarrow\theta_{\rvec,i}^\prime=\theta_{\rvec,i}-b_i\delta_\mu r_\mu
\end{equation}
in the free energy, and calculate

\begin{equation}
  \Upsilon_{\mu,(b_1,b_2)} = \frac{\partial^2 F[\theta^\prime]}{\partial\delta_\mu^2}\bigg|_{\delta_\mu=0}.
\end{equation}
Here, $b=(b_1, b_2)$ represents some combination of the phases $\theta_1$ and $\theta_2$,
$b_1\theta_1+b_2\theta_2$.
To probe the individual moduli, $b_i$ is chosen as $b_i=(1,0)$ or $b_i=(0,1)$.
The composite, phase sum variable is represented by the choice $b_i=(1,1)$, while
$b_i=(1,-1)$ is the phase difference.
Generally, for a two-component model, the helicity modulus can be written as the sum of two indivudual moduli, and a
cross term,\cite{Dahl2008_3,Herland2010}
\begin{equation}
  \Upsilon_{\mu,(b_1,b_2)} = b_1^2\Upsilon_{\mu,(1,0)}+b_2^2\Upsilon_{\mu,(0,1)} + 2b_1b_2\Upsilon_{\mu,12}.
  \label{eq:mixedmodulus}
\end{equation}
For the model considered in this paper, the individual helicity moduli can be written as

\begin{align}
  \langle\Upsilon_{\mu,i}\rangle=&\frac{1}{V}\bigg[\bigg\langle\sum_\rvec\psi_{\rvec_i}\psi_{\rvec+\hat{\boldsymbol{\mu}},i}\cos(\chi_{\rvec,i}^\mu)\bigg\rangle\nonumber\\
                                      &-\beta\bigg\langle\bigg(\sum_\rvec\psi_{\rvec,i}\psi_{\rvec+\hat{\boldsymbol{\mu}},i}\sin(\chi_{\rvec,i}^\mu)\bigg)^2\bigg\rangle\bigg],
\end{align}
while the mixed term has the form
\begin{align}
  \langle\Upsilon_{\mu,12}\rangle =
  -\beta\bigg\langle\bigg(&\sum_\rvec\left|\psi_{\rvec,1}\right|\left|\psi_{\rvec+\hat{\boldsymbol{\mu}},1}\right|\sin(\chi_{\rvec,1}^\mu)\bigg)\nonumber\\
  \bigg(&\sum_\rvec\left|\psi_{\rvec,2}\right|\left|\psi_{\rvec+\hat{\boldsymbol{\mu}},2}\right|\sin(\chi_{\rvec,2}^\mu)\bigg)\bigg\rangle.
  \label{eq:moduluscross}
\end{align}
We denote the helicity modulus of the phase sum $\Upsilon_{\mu,(1,1)}$ as $\Upsilon^+_\mu$ as a short
hand.

The structure factor, $S_i(\mathbf{q_\perp})$, can be used to determine the underlying symmetry
of the vortex lattice. Square and hexagonal vortex structures will manifest themselves as four or
six sharp Bragg peaks in reciprocal space. In a vortex liquid phase one expects a completely
isotropic structure factor.
The structure factor is defined as the Fourier transform of the longitudinally averaged
vortex density, $\langle n_i(\mathbf{r}_\perp)\rangle$, which is subsequently thermally
averaged,
\begin{equation}
S_i(\mathbf{q_\perp})=\frac{1}{L_xL_yf}\bigg\langle\bigg|\sum_{\rvec_\perp}
n_i(\rvec_\perp)\text{e}^{-\mathrm{i}\mathbf{r}_\perp\cdot\mathbf{q_\perp}}\bigg|\bigg\rangle.
\end{equation}
Here $n_i(\rvec_\perp)$ is the denisity of vortices of color $i$ averaged over the $z$-direction

\begin{equation}
  n_i(\rvec_\perp)=\frac{1}{L_z}\sum_z n_i(\rvec_\perp,z),
\label{eq:vdens}
\end{equation}
and $\rvec_\perp$ is $\rvec$ projected onto a layer of the system with a given $z$-coordinate. The
vortex density is calculated by traversing each plaquette of the lattice, adding the factor
$\chi_{i,\rvec}^\mu$ of each link. Each time we have to add (or subtract) a factor of $2\pi$ in
order to bring this sum back into the primary interval of $(-\pi,\pi]$ a vortex of color $i$
and charge +1(-1) is added to this plaquette.

In addition to the structure factor, we look at thermally averaged vortex densities, $\langle
n_i(\mathbf{r}_\perp)\rangle$, as well as thermally and longitudinally averaged amplitude
densities, $\langle\left|\psi_i\right|^2(\mathbf{r}_\perp)\rangle$, defined similarly to
Eq.~\eqref{eq:vdens},

\begin{equation}
\left|\psi_i\right|^2(\mathbf{r}_\perp)=\frac{1}{L_z}\sum_z
\left|\psi_i\right|^2(\rvec_\perp,z).
\label{eq:pdens}
\end{equation}
This provides an overview of the real space configuration of the system.

When including amplitude fluctuations, which, when the potential term is disregarded, are unbounded
from above, it is of great importance to make sure all energetically allowed configurations are
included. To this end, we measured the probability distribution of $\left|\psi_i\right|^2$,
$P(\left|\psi_i\right|^2)$ during the simulations by making a histogram of all field configurations
at each measure step, and normalizing its underlying area to unity in post-processing.

The uniform rotation applied to the condensates is implemented in the Landau gauge:

\begin{equation}
\mathbf{A}=(0,2\pi fx,0),
\end{equation}
where $f$ is the density of vortices in a single layer. Note that this implies a constraint $L f \in
(1, 2, 3, \ldots)$ due to the periodic boundary conditions.  When probing a first order melting
transition, it is important to choose a filling fraction large enough that an anomaly in the
specific heat is detectable. However, if the filling fraction is too large, one may transition
directly from a vortex liquid into a pinned solid, completely missing the {\em floating solid}
\/phase of interest. This scenario is characterized by a sharp jump in not only the longitudinal,
but also the transverse helicity modulus.\cite{Teitel94,Wheatley94} One must therefore chose $f$
small enough, to assure that the vortex line lattice is in a floating solid phase when it melts.

\section{Details of the Monte Carlo simulations}
\label{sec:MCdetails}

The simulations were performed using the Metropolis-Hastings
algorithm.\cite{Metropolis53,Hastings70} Phase angles were defined as $\theta\in(-\pi,\pi]$, and
amplitudes as $\left|\psi\right|^2\in(0,1+\delta\psi]$. The choice of $\delta\psi$ will be discussed
further, as it is important to ensure inclusion of the full spectrum of fluctuations. Both the
phases and the amplitudes were discretized to allow the use of tables for trigonometric and square root
functions in order to speed up computations. We typically simulated systems of size $L^3=64^3$, with
sizes up to $L^3=128^3$ used to resolve anomalies in the specific heat. We used $10^6$ Monte Carlo
sweeps per inverse temperature step, and up to $10^7$ close to the transition. $10^5$ additional
sweeps were typically used to thermalize the system. In the simulations, we examined time series of the internal
  energies taken during both the thermalization runs and the measurements runs to make sure the
simulation converged. One sweep consists of picking a new random configuration for each of the four
field variables separately in succession, at each lattice site. Measurements were usually performed
with a period of 100 sweeps, in order to avoid correlations. Ferrenberg-Swendsen multi-histogram
reweighting was used to improve statistics around simulated data points, and jackknife estimates of
the errors are used.

\begin{figure}
\centering
\includegraphics[width=\columnwidth]{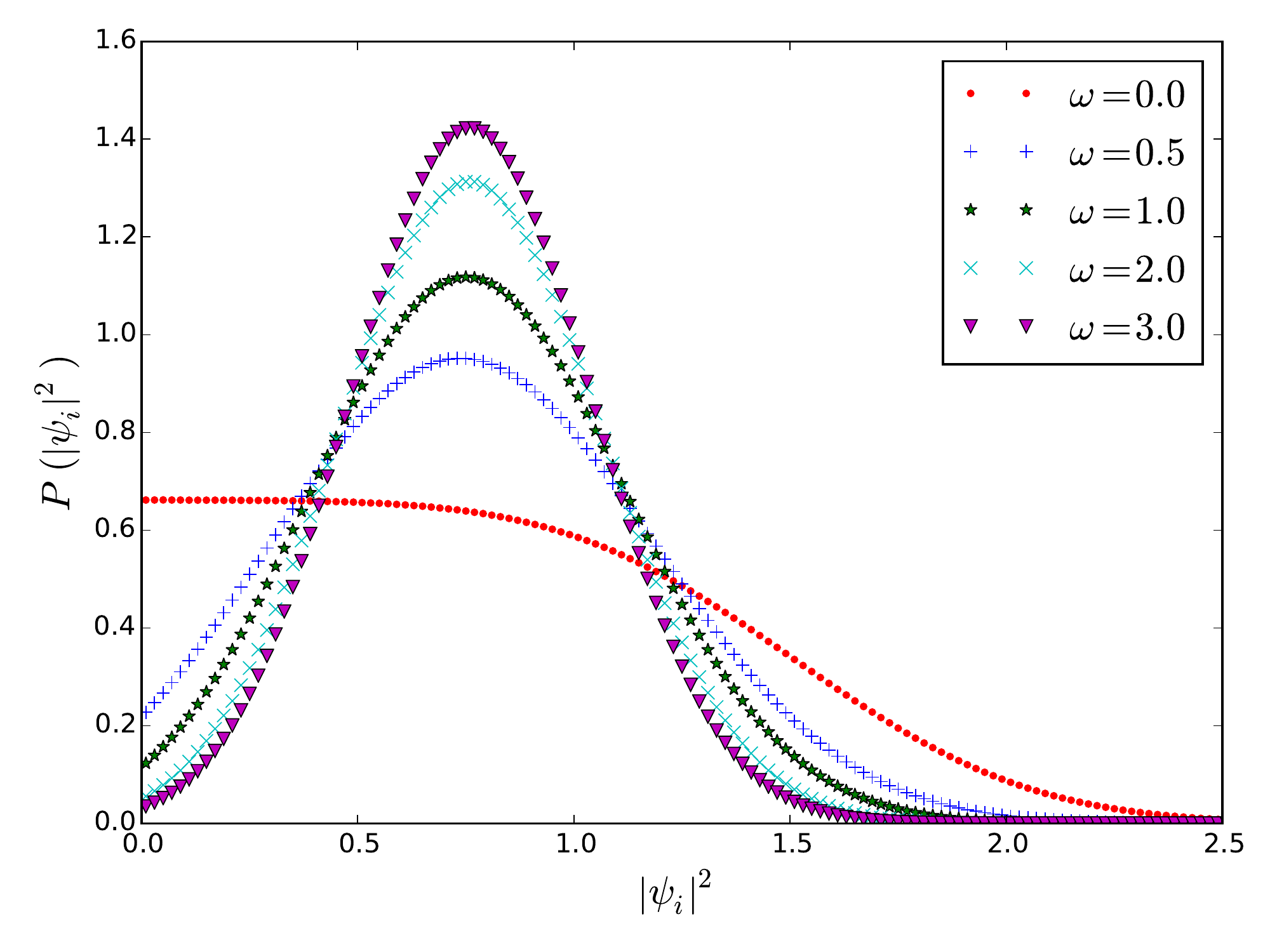}
\caption{(Color online) The probability distribution of the amplitudes, $\mathcal{P}(\abs{\psi_i}^2)$, for $N=2$, at inverse temperature $\beta=1.20$,
$f=1/32$, and $\eta=2$, with $\omega$ values from $0$ to $3$. The distribution is completely symmetric
in $i$.}
\label{fig:2compprob}
\end{figure}

Fig.~\ref{fig:2compprob} shows the probability distribution of the amplitudes,
$\mathcal{P}(\abs{\psi_i}^2)$. We get a peaked distribution for finite $\omega$. On the other hand,
when $\omega=0$, this is no longer the case. The distribution now approaches a uniform distribution
on the interval $(0,1]$. In this case the parameter $\eta$ serves to control the approach to
uniformity, $\eta\rightarrow\infty$ corresponding to the $CP^1$ limit.

With these initial simulation runs as a basis, we choose $\delta\psi$ appropriately in order to
capture the entire spectrum of fluctuations.

\section{Results of the Monte Carlo simulations}
\label{sec:results}

In this section, the $\eta-\omega$ phase diagram of ground states is explored by slow cooling and
examination of vortex and amplitude densities, as well as structure factors. In addition to the
expected hexagonal and square vortex ground states, several interesting regions of the parameter
space are investigated further. A special case between the square and the hexagonal region of the
phase diagram is discovered, where the lattice first forms a square structure, but thermally
reconstructs into a hexagonal lattice as the temperature is decreased further.  Furthermore, we
consider in detail the $\omega=0$ line in the phase diagram, where we discover additional vortex
fluctuation effects. For $\omega=0$, the system features an $\mathrm{SU(2)}$ symmetry.  An unusual
feature is an interesting state with global phase coherence, but  without a regular vortex lattice.
In this case ordinary vortices do not have topological character due to $\mathrm{SU}(2)$ symmetry.
Additionally, we obtain several interesting vortex structures characterized by dimer-like
configurations at lower temperatures. Here, we observe honeycomb lattices, or double-core lattices,
and stripe configurations, consistent with previous $T=0$ results.\cite{Kasamatsu2003}

We also examine the melting transitions of the square and hexagonal lattices with the full amplitude
distribution included, as well as the melting of the hexagonal lattice in a model with $N=1$ as a
benchmark of the method. To classify the transition, we look at thermal averages of the specific
heat, helicity moduli, and vortex structure factors. These results are presented in
Appendixes~\ref{sec:N1} and~\ref{sec:latticemelting}.

\subsection{The $\eta-\omega$ phase diagram}
\label{sec:nwphases}

\begin{figure}
  \subfloat[]{\includegraphics[width=\columnwidth]{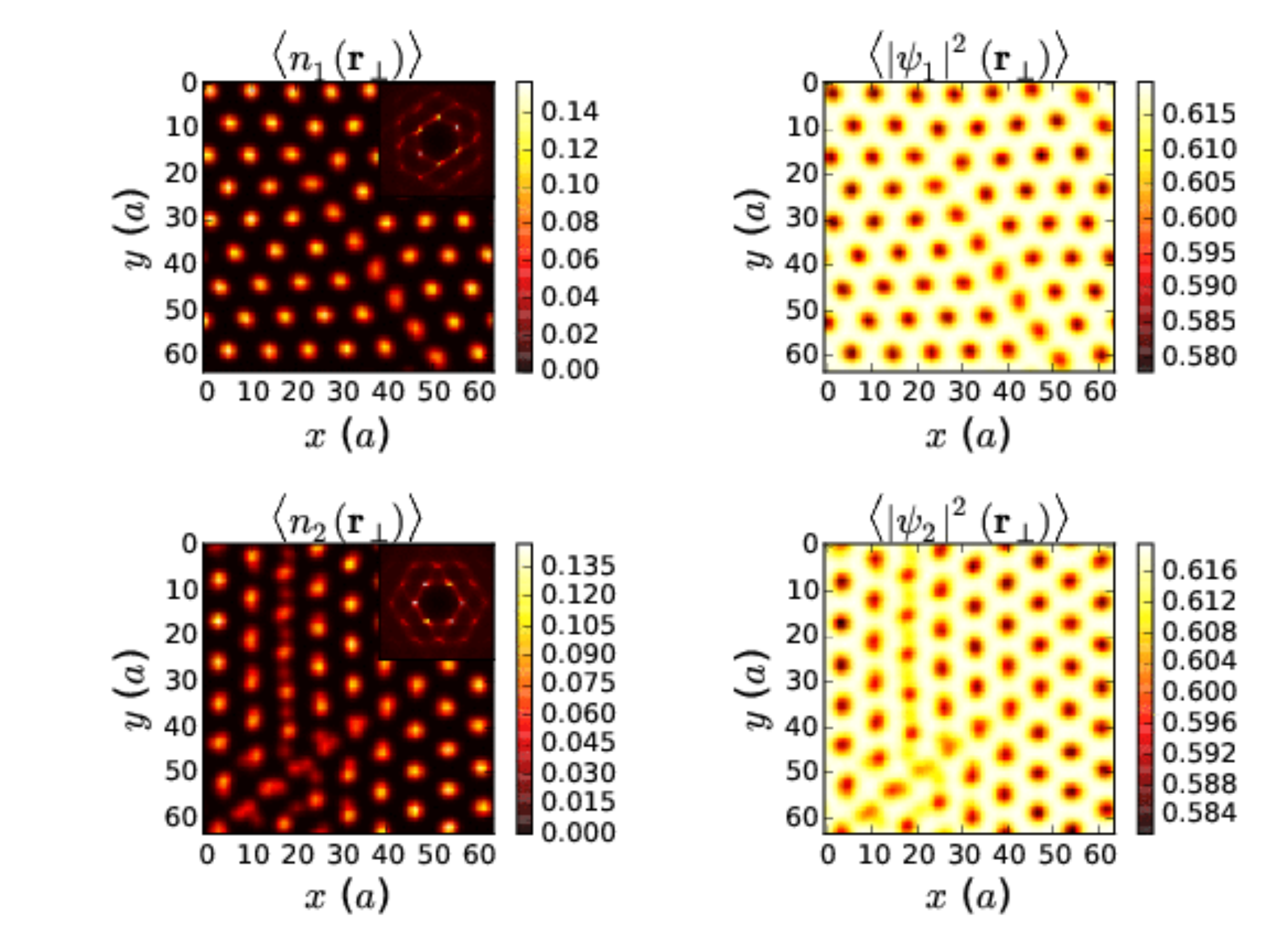}
  \label{fig:hexplot}}\\
  \subfloat[]{\includegraphics[width=\columnwidth]{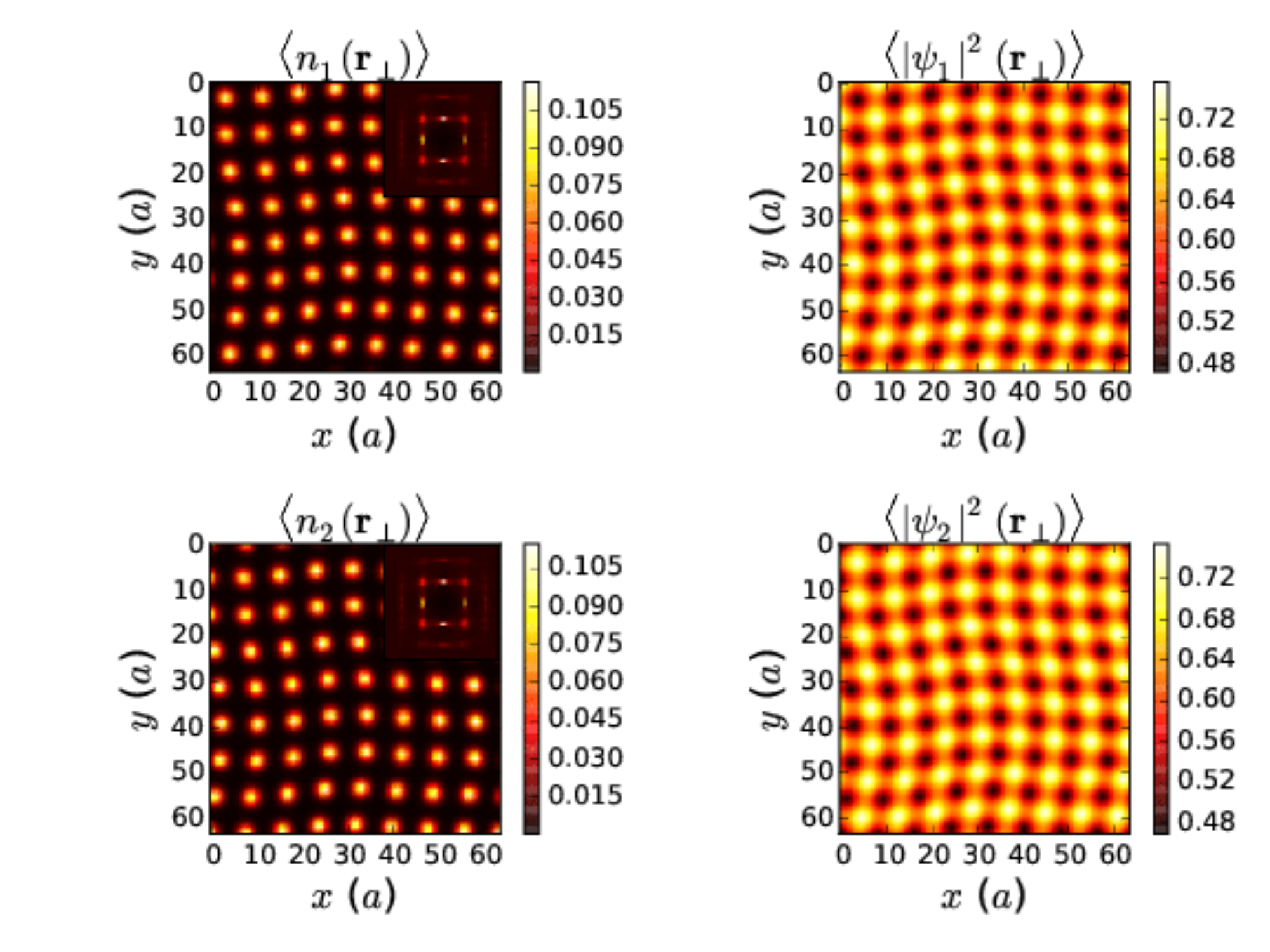}
  \label{fig:squareplot}}
  \caption{(Color online) Representative configurations of the two main ordered phases in the $\mathrm{U(1)}\times\mathrm{U(1)}$
    region. (a) shows a square structure at $(\eta,\omega)=(5.0,0.5)$, while (b)
    illustrates the hexagonal structure at $(\eta,\omega)=(5.0,5.0)$. Each subfigure shows vortex
    densities, $\langle n_i(\mathbf{r}_\perp\rangle$, in the left column, amplitude densities,
    $\langle\left|\psi_i\right|^2(\mathbf{r}_\perp\rangle$, in the right column, and structurue
  factors (insets) of each component as indicated. The induced vortex density and inverse
temperature are fixed to $f=1/64$ and $\beta=1.5$ in both subfigures.}
\label{fig:square_hex}
\end{figure}

Adding a second matter field and inter-component density-density interactions results in a
considerably richer set of ground states than in the single-component case. In the
absence of a fluctuating part of the rotational "gauge-field" there will be no gauge-field-mediated inter-component
current-current interactions. For $\eta - \omega < 0$, $(\eta,\omega) > 0)$ the effective
inter-component density-density coupling $\eta-\omega$ is negative and the ground state of each
color of condensate has a hexagonal symmetry, as shown in Fig.~\subref*{fig:hexplot}. If, on the other hand $\eta
-\omega > 0$, the inter-component coupling becomes positive. Now, for sufficiently large ratios
$\eta/\omega$, the vortices arrange themselves into two inter-penetrating square lattices, shown in
Fig.  \subref*{fig:squareplot}. The value of the ratio $\eta/\omega$ for which the lattice
reconstructs depends on the strength of the rotation, $f$.
If we neglect fluctuations, $\eta - \omega <
0$ is expected to result in a hexagonal lattice, while $\eta-\omega > 0$ leads to a square lattice
for sufficiently large $\eta/\omega$.

The physics of the reconstruction of the lattice can be explained by modulations of the amplitude
fields. The existence of static periodic amplitude-modulations (density-variations) is due to the
presence of vortices. Without vortices ($f=0$) and $\omega>0$, the ground state is one where both
amplitudes are equal and smooth. Vortices in one component tend to suppress locally the
corresponding amplitude, which in turn means that the term $\eta(|\psi_1|^2 + |\psi_2|^2-1)^2$
enhances the amplitude of the other component.  At small $\omega$, \textit{i.e.}\ large
$\eta-\omega$ there is a strong tendency to form a square density lattice due to this intercomponent
density-density interaction.  Conversely, if $\omega$ is large enough compared to $\eta$, the
density-density interaction  is not strong enough to overcome the isotropic current-current
interactions between same-species vortices.  In other words if the current-current interactions
dominate the interspecies density-density interactions, a hexagonal lattice is energetically
favoured over a square lattice, and vice versa. Note that similarly a square vortex lattice forms in
two-component London models with dissipationless drag when there are competing inter- and
intra-species current-current vortex interactions.\cite{Dahl2008,Dahl2008_2}

\begin{figure}
\centering
\subfloat{
\includegraphics[width=\columnwidth]{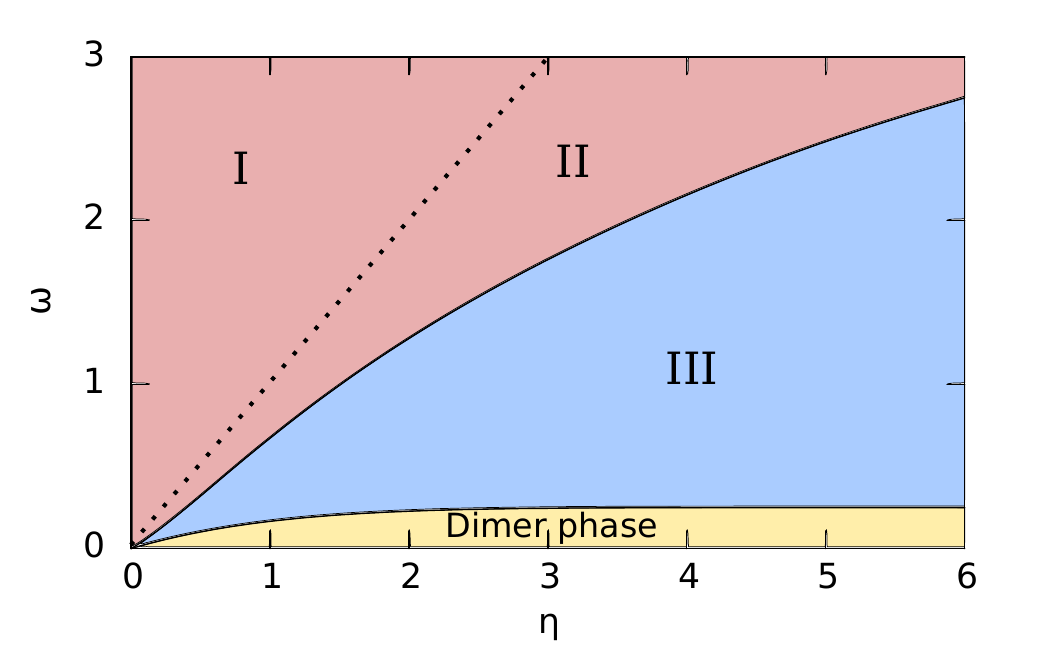}
\label{fig:phasesf_32}}

\subfloat{
\includegraphics[width=\columnwidth]{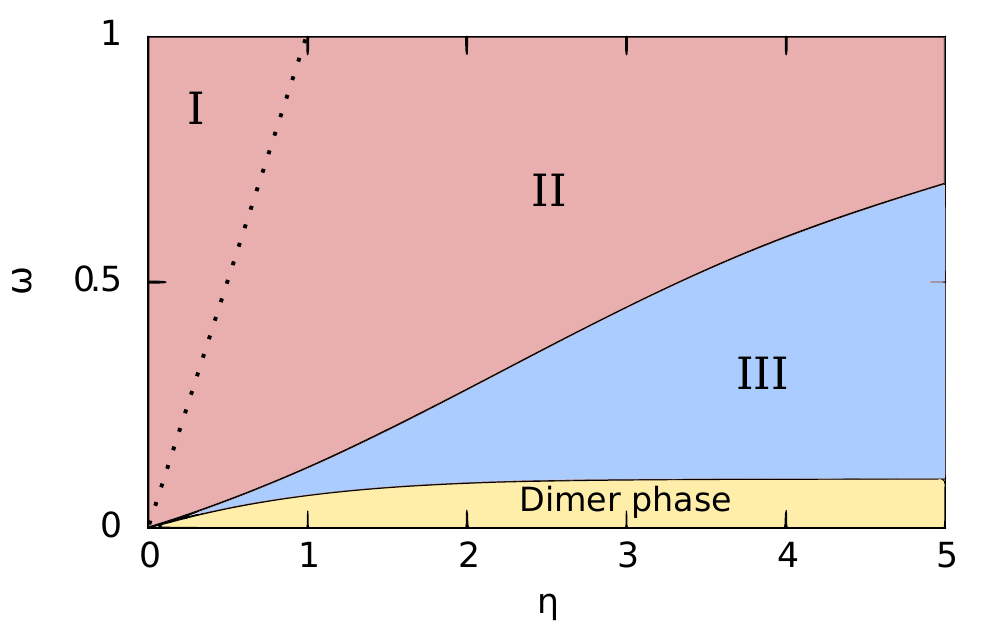}
\label{fig:phasesf_64}}
\caption{(Color online) The $\eta-\omega$ phase diagram of the ground states for $f=1/32$ (top) and
$f=1/64$ (bottom). The simulations were performed for a range of $(\eta,\omega)$ pairs to determine
the zero temperature ground state. Approximate demarcation lines for the phase boundaries separating
hexagonal lattices, square lattices, and dimerized phases, were drawn from these results (solid lines).
I denotes the  phase where the hexagonal vortex lattices in the two components are cocentric, II
denotes the case where the hexagonal lattices are intercalated, while III denotes the square lattice
phase. The dotted line is the line $\omega = \eta$ at which the intercomponent density-density
interaction $2(\eta-\omega) |\psi_1|^2 |\psi_2|^2$ changes sign. See also Figs.~\ref{fig:tableaux_l5}
and~\ref{fig:tableaux_l3} in Appendix~\ref{app:tableaux}.}
\end{figure}

Figures.~\subref*{fig:phasesf_32} and~\subref*{fig:phasesf_64}   show
the phase diagrams for filling fractions $f=1/32$ and $f=1/64$, respectively. The separation line
is approximate and drawn from several separate simulations.

To clarify what is going in Figs.~\subref*{fig:phasesf_32} and~\subref*{fig:phasesf_64}, we refer to
Figs.~\ref{fig:tableaux_l5} and~\ref{fig:tableaux_l3} in Appendix~\ref{app:tableaux}. Here, we show
tableaus to illustrate in more detail how the density and vortex lattices reconstruct at a
temperature well below any melting temperatures of the vortex (and density) lattices, as the
density-density interaction $2(\eta-\omega) |\psi_1|^2 |\psi_2|^2$ is varied. Specifically, we fix
the interaction parameter $\eta$, as well as the inverse temperature $\beta$ and filling fraction
$f$, while increasing the parameter $\omega$. This reduces the effective inter-component
density-density interaction which favors a square lattice, until the lattice reconstructs from
square to hexagonal symmetry.

When $\eta=\omega$,  it is seen from Eq.~\eqref{eq:inter_intra_g} that the two components of the order parameter
decouple. For $\omega<\eta$ the inter-component density-density interaction is repulsive, while it is
attractive for $\omega>\eta$. For $\omega  < \eta$, the vortex lattices (and the density lattices) are
intercalated, while for $\omega > \eta$  they are cocentric. In Figs.~\ref{fig:phasesf_32} and
\ref{fig:phasesf_64} we illustrate the demarcation line between the two situations as a dotted
line in the hexagonal phase.

Beyond the square and hexagonal lattices we also observe dimer configurations of vortices for $\omega=0$,
which will be discussed further below. The calculations are consistent with the ground states
obtained in Refs. \citenumns{Kasamatsu2003} and \citenumns{makoto2}.

\subsection{Thermally induced reconstruction of vortex lattices}

\begin{figure}
\centering
\subfloat[]{\includegraphics[width=0.5\columnwidth]{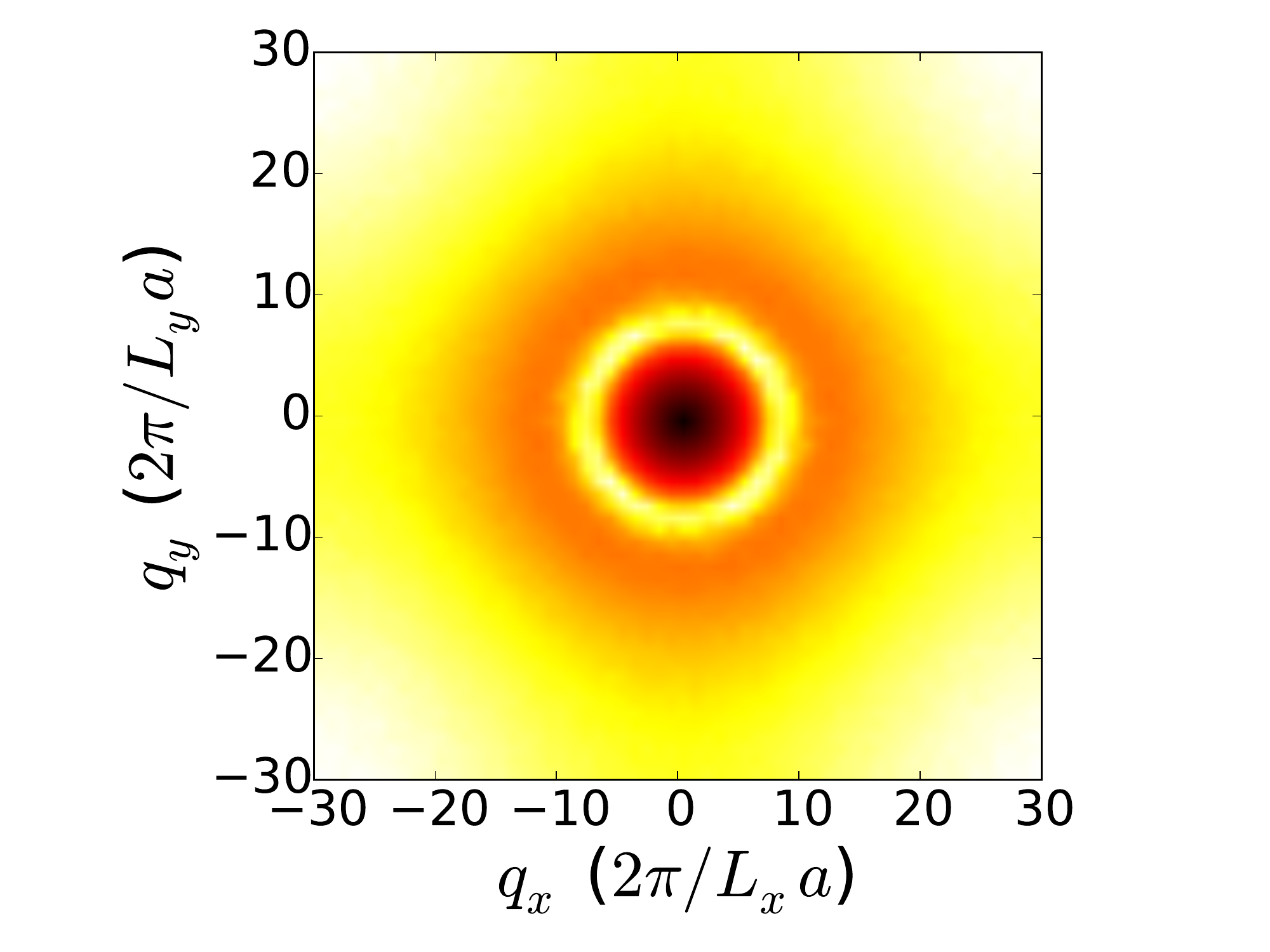}}
\subfloat[]{\includegraphics[width=0.5\columnwidth]{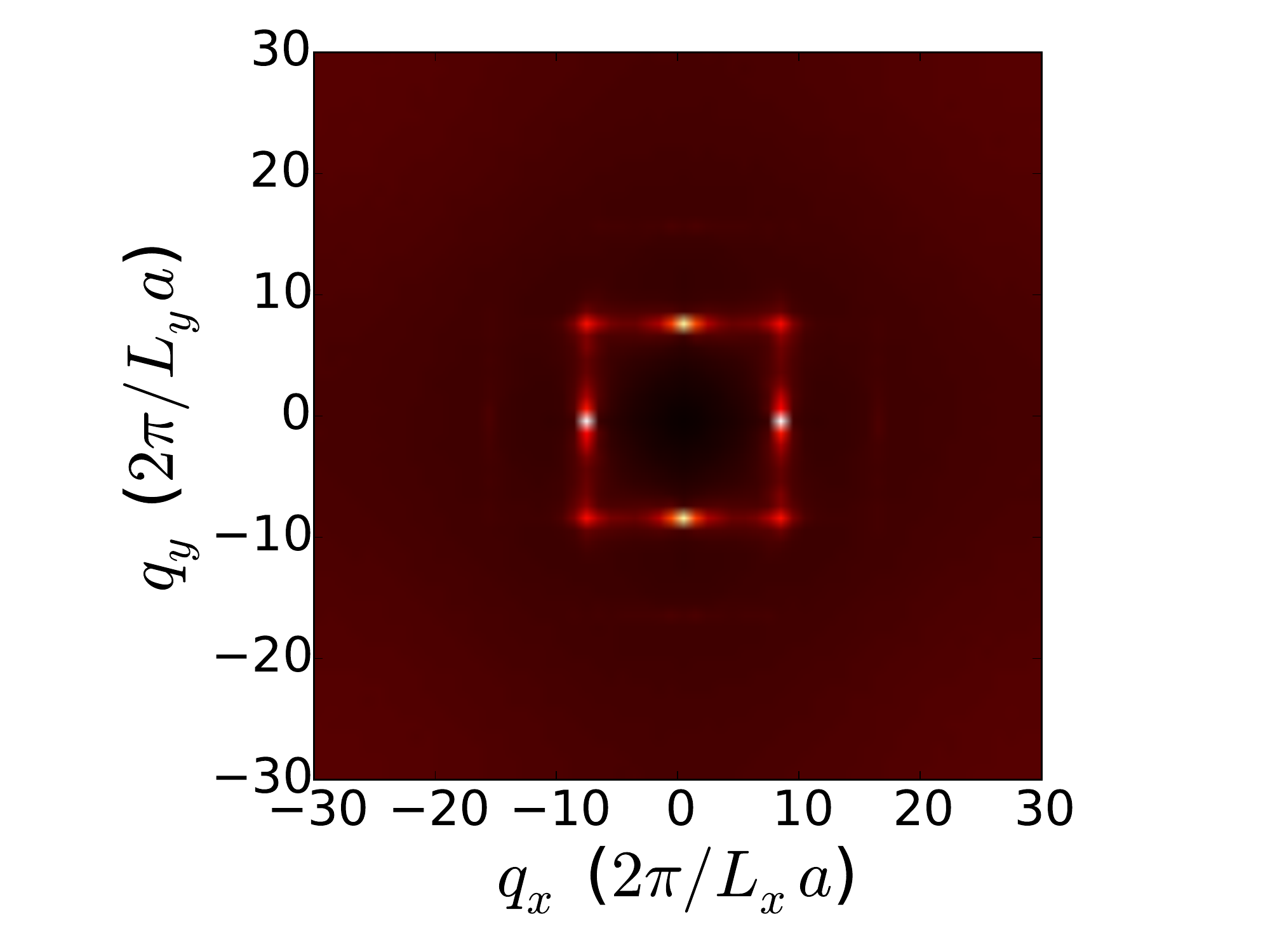}}\\
\subfloat[]{\includegraphics[width=0.5\columnwidth]{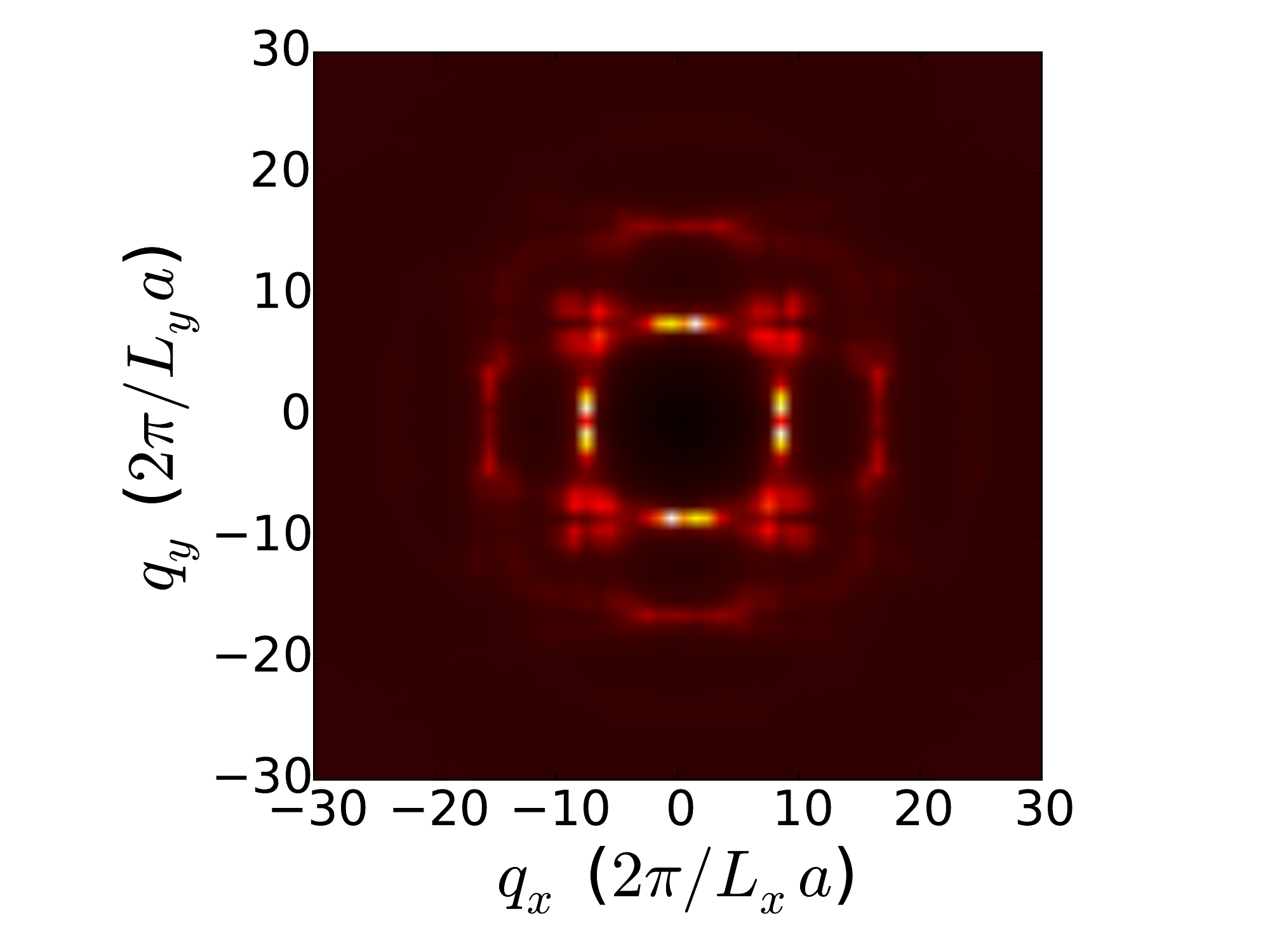}}
\subfloat[]{\includegraphics[width=0.5\columnwidth]{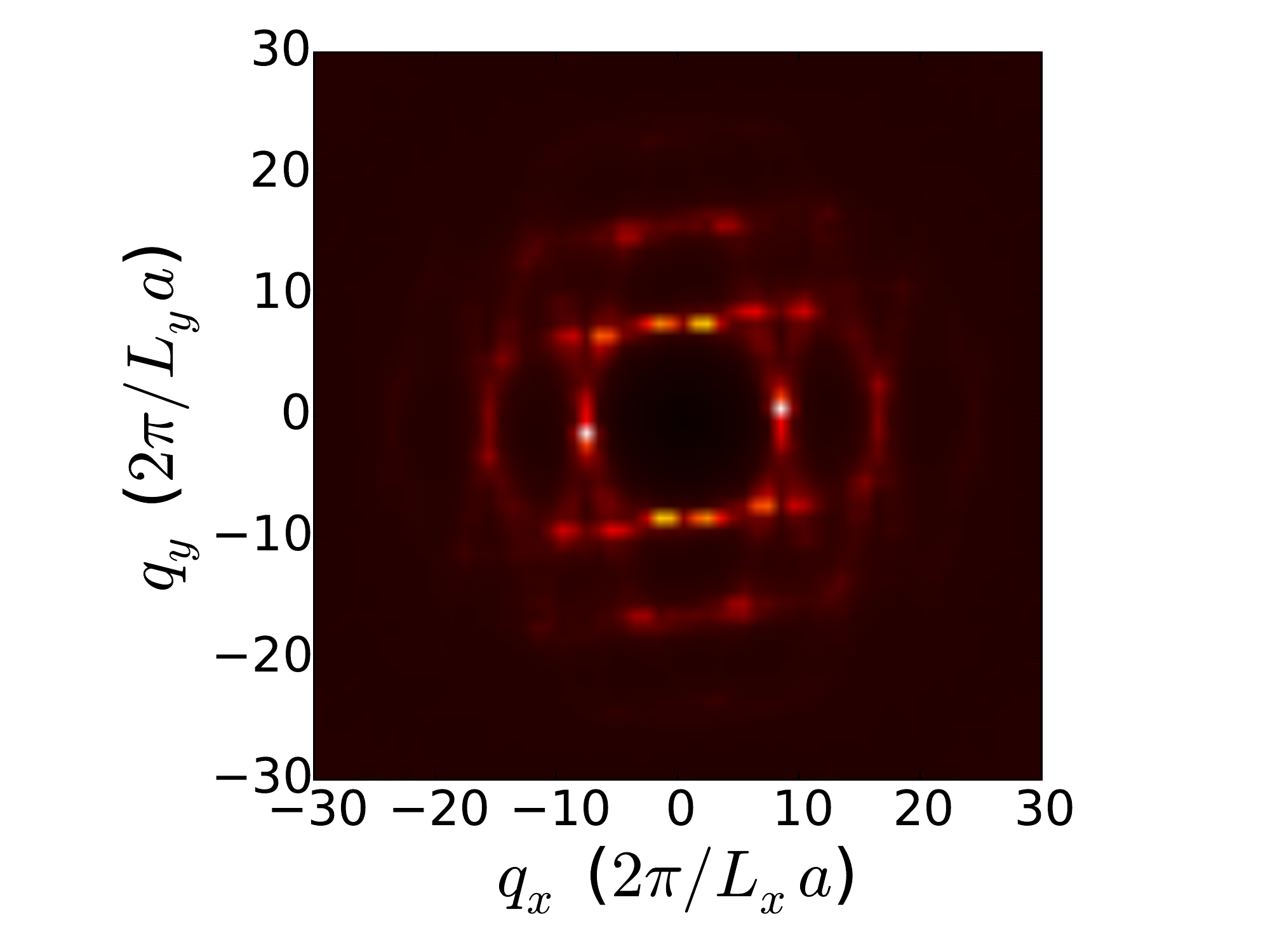}}\\
\subfloat[]{\includegraphics[width=0.5\columnwidth]{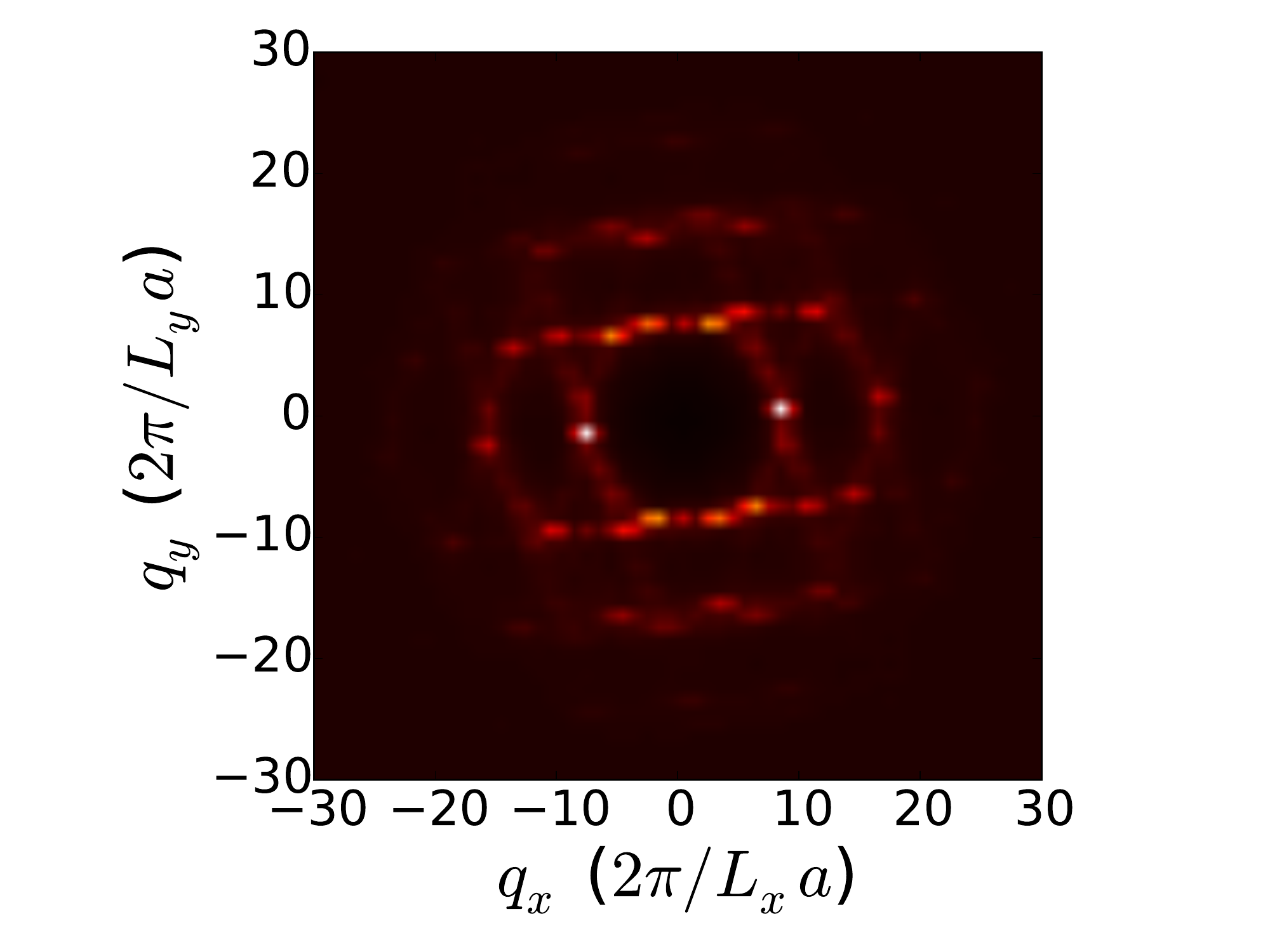}}
\subfloat[]{\includegraphics[width=0.5\columnwidth]{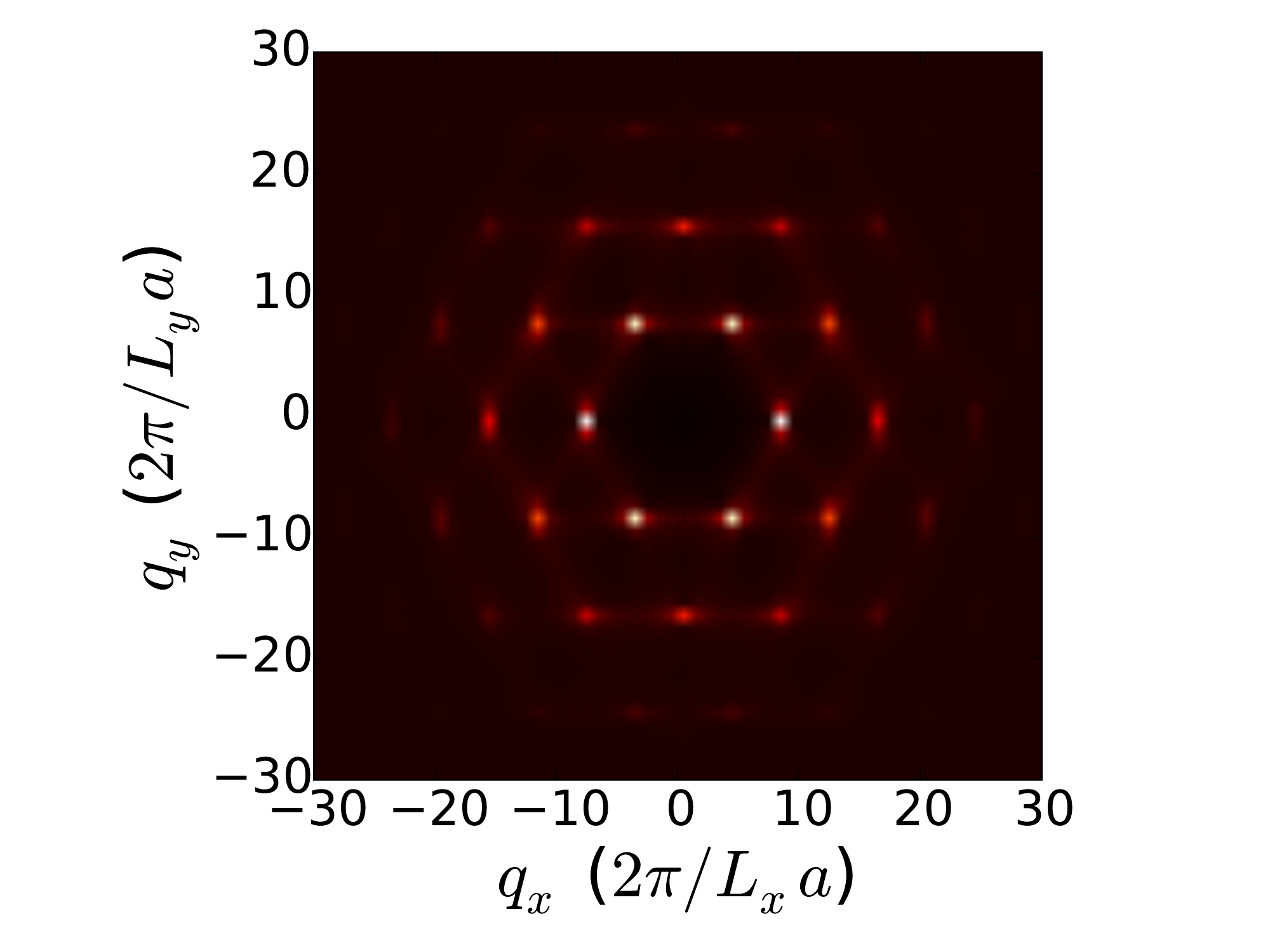}}
\caption{(Color online) Thermally induced reconstruction from a square vortex lattice in either of the components
    at $\eta=2,\omega=0.5$, to a hexagonal vortex lattice, as $\beta$ is increased. Here, $f=1/64$.
    (a)-(f) show inverse temperatures $\beta=\{0.80, 0.90, 1.20, 1.30, 1.34,
    1.38\}$, respectively
    Each subfigure shows $S_1(\mathbf{q}_\perp)$ only; $S_2(\mathbf{q}_\perp)$ is identical.
    The physical reason for the reconstruction originates with the inter-component density-density
    interaction term $2( \eta-\omega ) |\psi_1|^2 |\psi_2|^2)$, and is explained in detail in the
text.}
\label{fig:crossover}
\end{figure}

Now we move to discussion of the effects of thermal fluctuations in these systems.
Fig.~\ref{fig:crossover} shows the vortex-densities in component $1$ in reciprocal space, as $\beta$
is increased, \textit{i.e.},\ as temperature is reduced, in a temperature range below where the
lattice melts. The actual melting of the two component lattice is discussed in
Section~\ref{sec:latticemelting}. We fix the filling fraction $f=1/64$, as well as the interaction
parameters $\eta=2$ and $\omega=0.5$.

For the highest temperatures shown in Fig.~\ref{fig:crossover} the vortex lattice is square.
Upon cooling the system, the vortex lattice reconstructs into a hexagonal lattice, consistent with
the ground state phase-diagram of Fig.~\subref*{fig:phasesf_64}. The density-density interaction
term $2(\eta-\omega) |\psi_1|^2 |\psi_2|^2$ aids formation of a square lattice at higher
temperatures, while the current-current interactions drives the lattice towards a hexagonal
configuration when it is cooled further. This means that the free energy per vortex of the square
lattice, which is lower than that of the hexagonal lattice at $\beta=0.90$, has become larger than
that of the hexagonal lattice when $\beta=1.50$. This is essentially the combination of an energetic
and an entropic effect. We observe this reconstruction not too far away from the demarcation line
separating a square and a hexagonal vortex lattice. Deep inside the hexagonal phase in
Fig.~\subref*{fig:phasesf_64}, we observe direct vortex lattice melting from a hexagonal
lattice to a vortex liquid. We note that  intermediate entropically-stabilized vortex lattice
  phases were of a subject of interesting investigation in the different system of $U(1)\times U(1)$
  superconductors \cite{kivelson}, however the vortex interaction form is different in this case.

\subsection{$\mathrm{SU(2)}$ vortex states}
\label{sec:su2lattices}

The limit $\omega\rightarrow 0$ is quite different from the $\mathrm{U(1)} \times
\mathrm{U(1)}$-symmetric case $\omega \neq 0$. From Eq.~\eqref{eq:N2pot}, it is seen that the
Hamiltonian is invariant under $\mathrm{SU(2)}$ transformations of $\Psi$. Vortices, which are
topological in a $\mathrm{U(1)} \times \mathrm{U(1)}$ model, are no longer topological in the
$\mathrm{SU(2)}$ case. One may unwind a $2 \pi$ phase winding by entirely transferring density of
one component to the other, which may be done at zero energy cost.

\begin{figure}
\centering
\subfloat{
\includegraphics[width=\columnwidth]{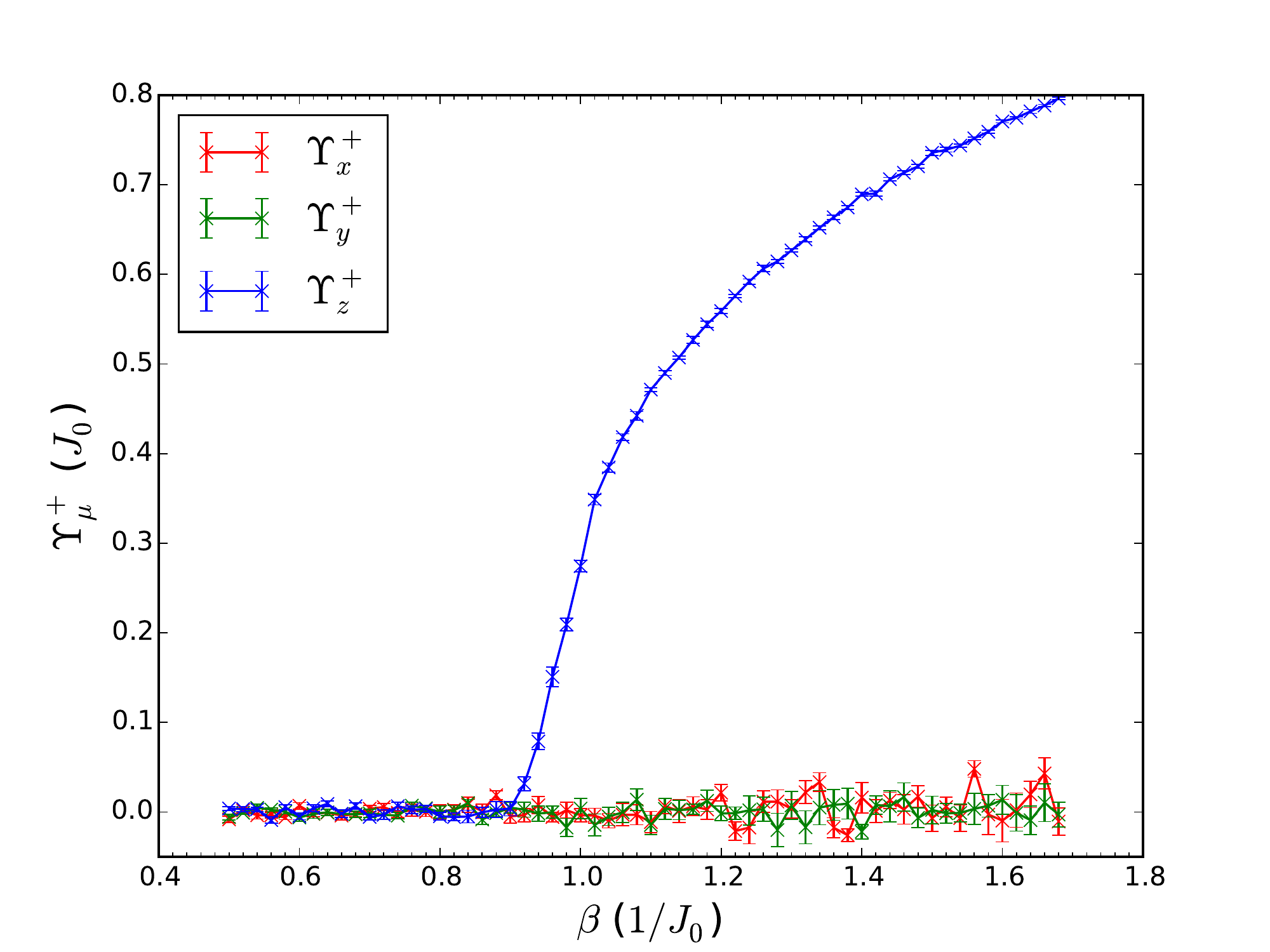}
\label{fig:su2heli}}\\
\subfloat{
\includegraphics[width=\columnwidth]{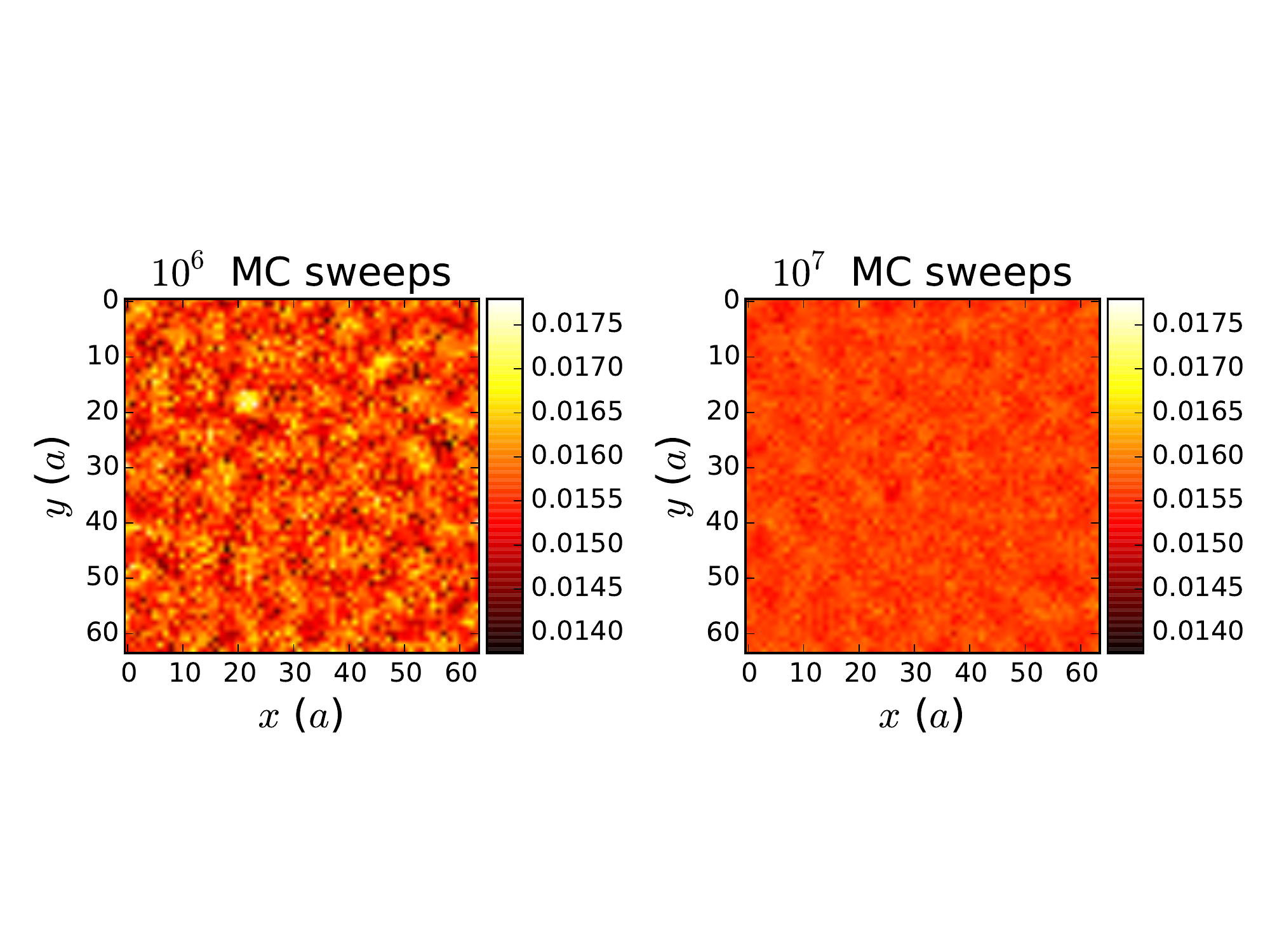}
\label{fig:su2vortsmooth}}
\caption{(Color online) Illustration of the observed state with coherence along the direction of the rotation axis
  without a regular vortex lattice, seen only with $\mathrm{SU(2)}$ symmetry. The parameters used
  are $\omega=0.0$, $\lambda=5$, and $f=1/64$. The top panel shows the helicity modulus of the phase
  sum, $\Upsilon^+_\mu$. The two bottom panels show the vortex densities, $n_1(\rvec_\perp)$, at
  $\beta=0.94$, where the $z$-directed modulus is clearly finite.  The bottom left and bottom right panels are
  taken from simulations using $10^6$ and $10^7$ Monte-Carlo sweeps, respectively. No apparent
  vortex line structure is seen here, and by increasing the number of Monte-Carlo sweeps the
  variations of the vortex density are smoothed out further.  Note how the
value of the average vortex density seems to converge towards $1/64$.}
\label{fig:su2obs}
\end{figure}

Fig.~\ref{fig:su2obs}  shows  one of the main results of our paper. These are simulations with
$\mathrm{SU(2)}$ symmetry, \textit{i.e.}, $\omega=0$, as well as $\eta=5.0$ and $f=1/64$. The top
panel show the phase stiffness associated with the phase sum, $\Upsilon^+_\mu$. This is the
physically relevant phase variable in this case, as it couples to the rotation.

We observe that the
stiffness along the $z$-direction becomes finite at an inverse temperature, $\beta\sim 0.9$. This is
what one would expect when a vortex lattice forms. However, the bottom panels, which shows the vortex
density of component $1$ at $\beta=0.94$, shows no apparent signs of vortex ordering. Hence, we have
an unusual situation. There is  a relatively large $\beta$-range where we have a finite $z$-directed
helicity modulus of the phase sum, but no apparent ordering of induced vortices.
A finite helicity modulus generally means that there are straight vortex lines with very little
transverse fluctuations threading the entire system along the direction in question. In the
$\mathrm{U(1)}$ picture this corresponds to a regular vortex lattice. For an
$\mathrm{SU(2)}$-condensate, this is no longer the case. Large relative amplitude fluctuations can
occur since they have zero energy cost in the ground state as the energy is no longer minimized by
a preferential value of $\left|\psi_1\right|^2-\left|\psi_2\right|^2$. This results in many
(nearly) degenerate vortex states between which the system can fluctuate, thus greatly simplifying
the effort of moving an entire, almost straight, vortex line. We are left with a phase where we have
coherence along the $z$-direction, but no regular vortex lattice appears in thermal averages. Nearly
straight vortex lines will shift between a large number of degenerate, or nearly degenerate, states
at a time scale shorter than a typical Monte-Carlo run.

The bottom panels of Fig.~\ref{fig:su2obs} show some
inhomogeneities of the vortex densities, exemplifying that this is not an ordinary vortex liquid
with segments of vortex lines executing transverse meanderings along their direction, which would
yield zero helicity modulus along the direction of the field-induced vortices.  Rather, what we have
is a superposition of many lattice-like states of nearly straight vortex lines, where the
fluctuations are largely collective excitations of entire nearly straight lines, rather than
fluctuations of smaller segments of lines.

We emphasize again that these collective excitations
originate with large amplitude fluctuations due to the $SU(2)$-softness of the amplitudes of the
components of the superfluid order parameter, rather than with phase fluctuations. Increasing the
number of Monte-Carlo sweeps by an order of magnitude smooths these variations out (without
noticably altering the value of $\Upsilon^+_z$), as seen in the bottom right panel of
Fig.~\ref{fig:su2obs}. Note how the average value of the vortex density seems to
converge towards $1/64$. This is what we expect for a vortex lattice or liquid in a
  $\mathrm{U}(1)\times\mathrm{U}(1)$ symmetric model, as the density of thermal
vortices will average to zero, and $f$ is the average flux density per plaquette.

\begin{figure}
\centering
\subfloat[]{
  \includegraphics[width=\columnwidth]{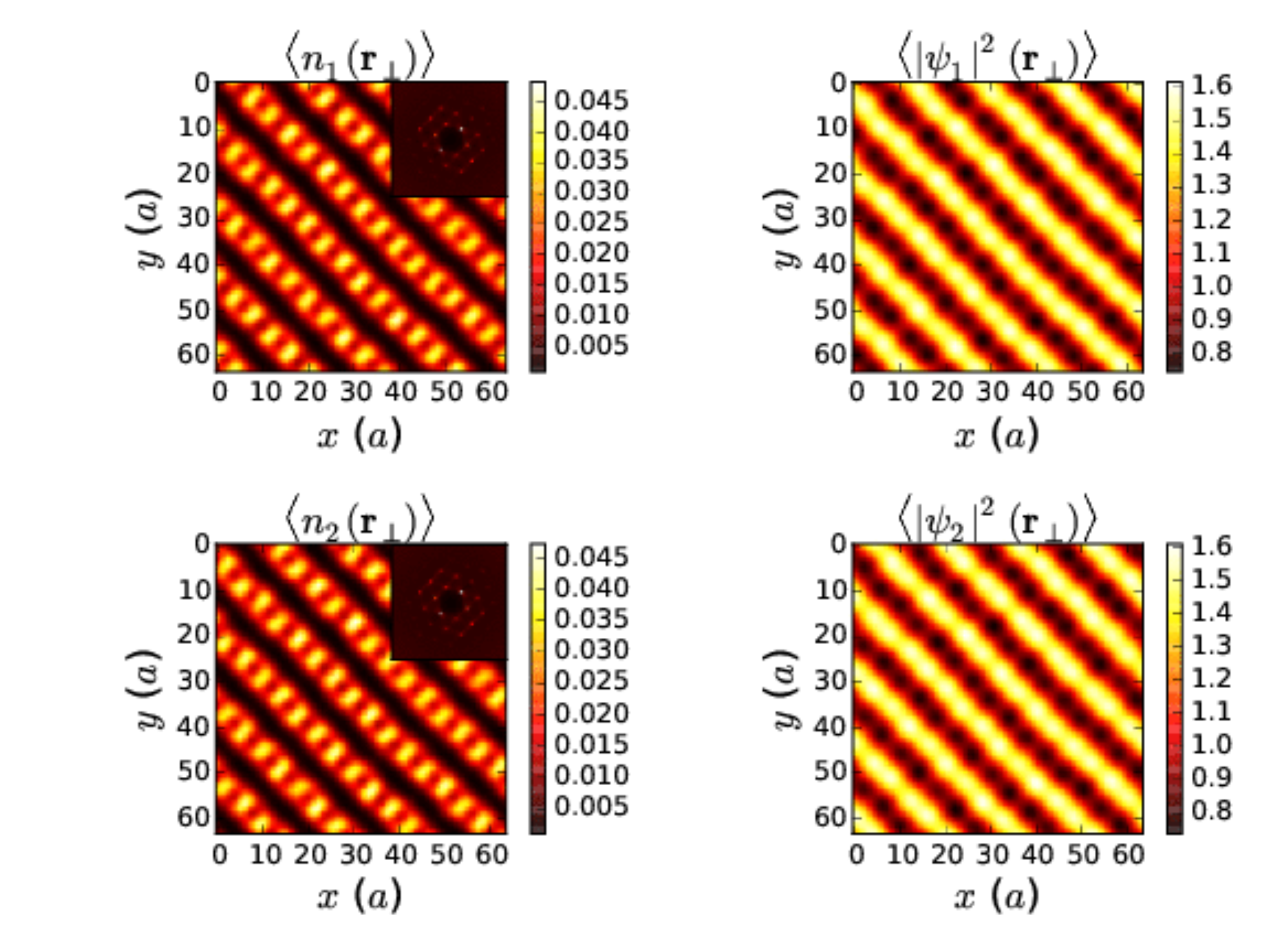}
\label{fig:su2vortices1}}\\
\subfloat[]{
  \includegraphics[width=\columnwidth]{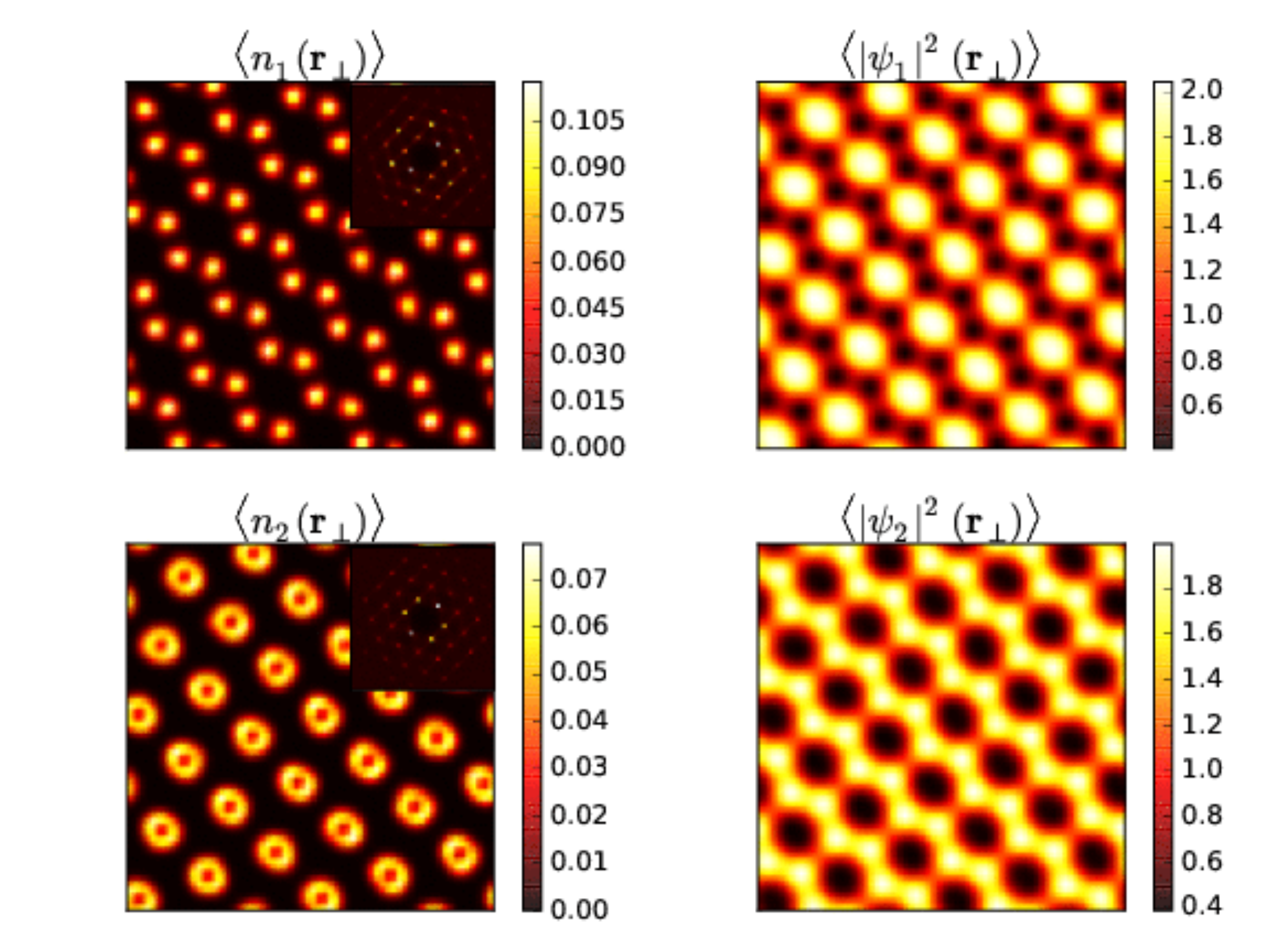}
\label{fig:su2vortices2}}
\caption{(Color online) Two examples of $\mathrm{SU(2)}$ vortex configurations from a single simulation, for two different inverse
  temperatures. The parameters $\eta$, $f$, and $\omega$ are fixed in each subfigure, at $\eta=1.0$,
  $\omega=0.0$, and $f=1/64$. (a) shows $\beta=0.84$, while (b) shows
  $\beta=1.50$. Each subfigure shows vortex densities, $\langle n_i(\mathbf{r}_\perp\rangle$, in
  the left column, amplitude densities, $\langle\left|\psi_i\right|^2(\mathbf{r}_\perp\rangle$, in
  the right column,  and structure factors (insets) of each component as indicated. This
    illustrates the degeneracy of the vortex line lattice in the isotropic limit, as the
    configurations evolve when $\beta$ is varied. See appendix~\ref{app:tableaux} for more
  details.}
\label{fig:su2vortices}
\end{figure}

As the system is cooled further, the movements of large vortex lines cease, and a regular vortex
lattice appears. However, degeneracy must still be present, as the exact pattern formed by the
lattice is distinctively different between simulations (keeping all parameters equal). The lattice
also has a tendency to shift between configurations as the temperature is varied, below the
temperature of initial vortex lattice formation.  We observe two distinct classes of vortex states,
illustrated in Fig.~\ref{fig:su2vortices}. The two are stripes (Fig.~\subref*{fig:su2vortices1}) and
honeycomb lattices (Fig.~\subref*{fig:su2vortices2}), both of which are seen in
Ref.~\citenumns{Kasamatsu2003}. Note that these vortex densities are taken from a single simulation,
after the lattice has formed. {Within the accuracy of our simulations the obtained states are
not metastable. The evidence of this is obtained by performing several independent runs from
different initial configurations.} Again, we refer to Appendix~\ref{app:tableaux}, where
Fig.~\ref{fig:tableaux_su2} illustrates the degeneracy in the vortex line lattices obtained in the
isotropic limit in further detail.

\section{Experimental considerations}
\label{sec:exp}

Hexagonal and square lattices have already been observed in binary condensates of rubidium.~\cite{Cornell2004} However,
an $\mathrm{SU(2)}$ condensate has not been realized experimentally. In this section, we briefly outline under what
circumstances an observation of an  $\mathrm{SU(2)}$ vortex state may be feasible.

In order to experimentally realize $\mathrm{SU(2)}$ conditions, one requires a two-component BEC,
where both intra- and inter-component interactions are equal. As we have seen, the $\mathrm{SU(2)}$
physics crucially depends on this, since even minor deviations from this condition immediately
yield $U(1) \times U(1)$ physics. This corresponds to $\omega=0$ in our parametrization.
Intra- and inter-component density-density interactions are given in terms of scattering lengths. While tuning
of these in an experiment is possible with Feshbach resonances, it may still be a challenge to tune
two scattering lengths independently to be equal to a third, to arrive at the $\mathrm{SU(2)}$ point.
From what is known for scattering lengths of real systems, it appears that a mixture of two species
of the same atom, but in different hyperfine states, lends itself more readily to a
realization of  an $\mathrm{SU(2)}$ condensate than a mixture of different atoms or a mixture of
different isotopes of the same atom. This is so, since in the former case, the relevant scattering
lengths typically {\it a priori} \/are much more similar to each other than what they are in more
heterogeneous mixtures.

One promising candidate therefore appears to be a condensate of $^{87}\text{Rb}$ prepared in the two
hyperfine states $|F=1, m_f=1\rangle\equiv|1\rangle$ and $|F=2, m_f=-1\rangle\equiv|2\rangle$. In
this system, the three relevant $s$-wave scattering lengths already have values close to the point of
interest, $a_{11} = 100.4 a_B$, $a_{22} = 95.00a_B$, and $a_{12} = 97.66a_B$, where $a_B$ is the
Bohr radius.~\cite{Cornell98,Merril2007} Reference \citenumns{Tojo2010} reports on a \textit{magnetic}
Feshbach resonance at a field of approximately $9.1$ Gauss, where control of $a_{12}$ of the order
of $10a_B$ is possible. Additionally, Reference \citenumns{Thalhammer2005} reports on an \textit{optical}
Feshbach resonance of the state $|F=1, m_F=-1\rangle$, able to tune the intracomponent scattering
length, using two Raman lasers, with detuning parameters approximately given by  $\Delta_1=2\pi\times75$MHz and
$\Delta_2=2\pi\times20$MHz. Here, varying $\Delta_2$ tunes the value of the scattering length around
the Feshbach resonance, while varying $\Delta_1$ changes the width. Hence, greater control of the
resonance is possible with an optical Feshbach-resonance compared to a magnetic one. Presumably, there
should exist optical Feshbach resonances able to tune the scattering length of either the $|1\rangle$
or the $|2\rangle$ state, for instance the one reported to exist at $1007$G for the $|1\rangle$
state.~\cite{Volz2003} This resonance should be far enough away from the inter-component resonance
at $9.1$G to not cause any interference.

This suggests one possible setup. Namely, prepare a two-component condensate of $^{87}$Rb in the
$|1\rangle$ and $|2\rangle$ states under rotation, and tune $a_{12}$ to $a_{22}$ using a magnetic
field. Then, tune $a_{11}$ to the same value using optical techniques, while taking time-of-flight
images of the condensate. The prediction is that as the system is tuned through the optical Feshbach
resonance, one should observe a hexagonal composite vortex lattice at sub-resonance frequencies, the
non unique vortex ordering pattern, discussed above, at a frequency  where all scattering lengths are equal, close to the
optical Feshbach resonance, and finally the reappearance of a hexagonal vortex lattice at
frequencies above the frequency where all lengths are equal Fig.~\ref{fig:su2vortices}. The
observation of a featureless rotating condensate would be a direct manifestation of the loss of
topological character of $U(1)$-vortices in the $\mathrm{SU(2)}$-symmetric case.
It would be interesting to study the dynamics of the vortex lattice in this case with methods like those used in
\cite{Freilich03092010}. For other discussions of SU($N$) models in cold atoms see Refs.
\citenumns{2014arXiv1403.2792C} and \citenumns{2010NatPh...6..289G}.

In actual experiments, a magnetic trap is used to confine the condensate in a given lateral region.
The effect of this on thermal fluctuations in vortex matter has been studied in detail in previous
theoretical works  for the one-component case, without amplitude fluctuations \cite{PhysRevLett.97.170403,PhysRevA.77.043605}.
The effect of the trap is to yield a maximum overall condensate density at the center of the trap, while depleting it
towards the edge of the trap. As a result, the lattice melts more easily near the edge of the trap. As can be inferred from the work
on single-component melting \cite{PhysRevLett.97.170403,PhysRevA.77.043605}, the results of the present paper, where no
inhomogeneity due to a magnetic trap has been accounted for, is therefore most relevant to the region close to the center
of the trap.

\section{Conclusions}
\label{sec:conclusions}

In this paper, we have investigated a two-component $\mathrm{U}(1)\times\mathrm{U}(1)$  and
$\mathrm{SU}(2)$ Bose-Einstein
condensate with density-density interaction under rotation at finite temperature, thereby extending
previous works which calculated the zero-temperature ground state numerically. In the
$\mathrm{U}(1)\times \mathrm{U}(1)$ case we
report that thermal fluctuations can lead to a phase transition between hexagonal and square vortex
lattices with increased temperature.

In the isotropic, $\mathrm{SU(2)}$, limit, we have observed an intermediate state of global phase
coherence without an accompanying vortex lattice in the thermally averaged measurements. In
  addition, we observe a variety of dimerized vortex states, such as dimerized stripes and
  honeycomb-like lattices, which exist for a wide range of temperature. These lattices could be
  observed in binary Bose-Einstein condensates in two separate hyperfine states, by precisely tuning the inter-
and intra-component scattering lengths to the $\mathrm{SU(2)}$ point through the use of Feshbach
resonances.

\begin{acknowledgments}
  We thank Erich Mueller for useful discussions.  P.~N.~G. thanks NTNU and the Norwegian Research
  Council for financial support. E.~B. was supported by by the Knut and Alice Wallenberg Foundation
  through a Royal Swedish Academy of Sciences Fellowship, by the Swedish Research Council Grants No.
  642-2013-7837 and No. 325-2009-7664, and by the National Science Foundation under the CAREER Award
No. DMR-0955902, A.~S. was supported by the Research Council of Norway, through Grants No. 205591/V20 and
No. 216700/F20.  This work was also supported through the Norwegian consortium for high-performance
computing (NOTUR).
\end{acknowledgments}

\appendix

\section{Rewriting the general Hamiltionian}
\label{app:rewrite}
Here we present the details of rewriting Eq.~\ref{eq:genH} into Eq.\ref{eq:Ham}, which is more
suited for our purposes. We repeat the starting point here for convenience.
\begin{align}
  H=\int
  d^3r\Bigg[&\sum_{i=1}^N\sum_{\mu=1}^3\frac{\hbar^2}{2m_i}\left|(\partial_\mu-\mathrm{i}\frac{2\pi}{\Phi_0}A_\mu^\prime)\psi_i^\prime\right|^2\nonumber\\
                            &+\sum_i^N\alpha_i^\prime\abs{\psi_i^\prime}^2+
\sum_{i,j=1}^Ng_{ij}^\prime\abs{\psi_i^\prime}^2\abs{\psi_j^\prime}^2\Bigg]
\end{align}
{First, we scale the field variables and Ginzburg-Landau parameters, to obtain some dimensionless
quantities.
\begin{align}
  \alpha_i^\prime &= \alpha_0\alpha_i,\\
  g_{ij}^\prime &= g_0g_{ij},\\
  \abs{\psi_i^\prime} &= \sqrt{\frac{\alpha_0}{g_0}}\abs{\psi_i}.
\end{align}
This gives us the Hamiltonian,
\begin{align}
  H=\frac{\alpha_0^2}{g_0}\int
  d^3r\Bigg[&\sum_{i=1}^N\sum_{\mu=1}^3\frac{\hbar^2}{2m_i\alpha_0}\left|(\partial_\mu-\mathrm{i}\frac{\Phi_0}{2\pi}A^\prime_\mu)\psi_i\right|^2\nonumber\\
            &+\sum_i^N\alpha_i\abs{\psi_i}^2+
\sum_{i,j=1}^Ng_{ij}\abs{\psi_i}^2\abs{\psi_j}^2\Bigg],
\end{align}
which on the lattice reads
\begin{align}
  H=\frac{\alpha_0^2a^3}{g_0}&\sum_\rvec\Bigg[\sum_{i=1}^N\sum_{\mu=1}^3\frac{\hbar^2}{m_i\alpha_0a^2}\nonumber\\
  \times\bigg(\big|\psi_{\rvec,i}\big|^2&-\big|\psi_{\rvec+\hat{\boldsymbol{\mu}},i}\big|\big|\psi_{\rvec,i}\big|\cos(\theta_{\rvec+\hat{\boldsymbol{\mu}},i}-\theta_{\rvec,i}-A_{\mu,\rvec})\bigg)\nonumber\\
                                   +&\sum_i^N\alpha_i\abs{\psi_{\rvec,i}}^2+
\sum_{i,j=1}^Ng_{ij}\abs{\psi_{\rvec,i}}^2\abs{\psi_{\rvec,j}}^2\Bigg],
  \label{eq:Hamoriglattice}
\end{align}
where $a$ is the lattice constant, and we have introduced
\begin{equation}
  A_\mu = \frac{2\pi}{\Phi_0} aA_\mu^\prime.
\end{equation}
}

Next, we specialize to the case $N=2$, $\alpha_1=\alpha_2$, $g_{11}=g_{22}\equiv g$, and $m_1=m_2$, and
define $a^2$ to be equal to $\hbar^2/m\alpha_0$, which sets our length scale.  Note that it should not
be confused with the coherence length in the multi-component case without intercomponent density-density interaction.
For the definition of coherence lengths in the presence of multiple components and inter-component
density-density interactions, see Refs. \citenumns{2011PhRvB..83q4509C} and
\citenumns{PhysRevB.90.064509}. The energy scale is defined as  $J_0$ as follows
\begin{equation}
  J_0=\frac{\alpha_0^2a^3}{g_0}.
\end{equation}
The coupling parameters $\eta$ and $\omega$ used in this paper are defined by comparing the potential
term of Eq.~\eqref{eq:Hamoriglattice} to the form where the soft constraints $\abs{\psi_1}^2+\abs{\psi_2}^2=1$ and
$\abs{\psi_1}^2-\abs{\psi_2}^2 = 0$ are implemented. Thus, we have
\begin{equation}
V(\Psi)=\eta(\left|\psi_1\right|^2+\left|\psi_2\right|^2-1)^2+\omega(\left|\psi_1\right|^2-\left|\psi_2\right|^2)^2,
\end{equation}
with

\begin{align}
  \eta &= -\frac{\alpha}{2}-\frac{3}{2}.\\
  \omega &= \frac{g-g_{12}}{2}
\end{align}

The lattice version of the Hamiltionian reads
\begin{align}
  H=&\sum_{\substack{\rvec,\hat{\boldsymbol{\mu}}\\i}}\big|\psi_{\rvec+\hat{\boldsymbol{\mu}},i}\big|\big|\psi_{\rvec,i}\big|
                                   \bigg(\cos(\theta_{\rvec+\hat{\boldsymbol{\mu}},i}-\theta_{\rvec,i}-A_{\mu,\rvec})\bigg)\nonumber\\
  +&\sum_\rvec \eta(\left|\psi_1\right|^2+\left|\psi_2\right|^2-1)^2\nonumber\\
                                 +&\sum_\rvec\omega(\left|\psi_1\right|^2-\left|\psi_2\right|^2)^2.
\end{align}
This model will then have the following continuum form,
\begin{equation}
  H = \int
  d^3r\Bigg[\sum_i^N\frac{1}{2}\left|(\partial_\mu-\mathrm{i}A_\mu)\psi_i\right|^2+V(\Psi)\Bigg].
\end{equation}

\section{First order lattice melting for $N=1$ reconsidered}
\label{sec:N1}
As a benchmark on simulations with amplitude fluctuations included, we verify the well-established first-order
melting transition on this model with only a single component of the order-parameter field, in the presence of amplitude
fluctuations. The added feature of the computation is that the complete amplitude-distribution function was utilized,
through the methods described in Section~\ref{sec:MCdetails}. In this case, the term in the potential proportional to
$\omega$ in Eq. \ref{eq:N2pot} is absent, and the potential reduces to
\begin{equation}
V(\Psi)=\eta(\left|\Psi\right|^2-1)^2.
\end{equation}
With amplitude fluctuations neglected, this model reduces to the much studied uniformly frustrated
3D$XY$ model, with well known results as mentioned in the Introduction of the paper. The model features
a first-order phase transition manifested as a melting of the frustration-induced hexagonal lattice of
vortices.\cite{Teitel91,Teitel97,Hu97,Nguyen98_1,Nguyen98_2,Stroud98,Nguyen99,Chin99,Nguyen_EPL99}
The fluctuations responsible for driving this transition are massless transverse phase fluctuations
of the order parameter.

\begin{figure}
\centering
\subfloat{
\includegraphics[width=\columnwidth]{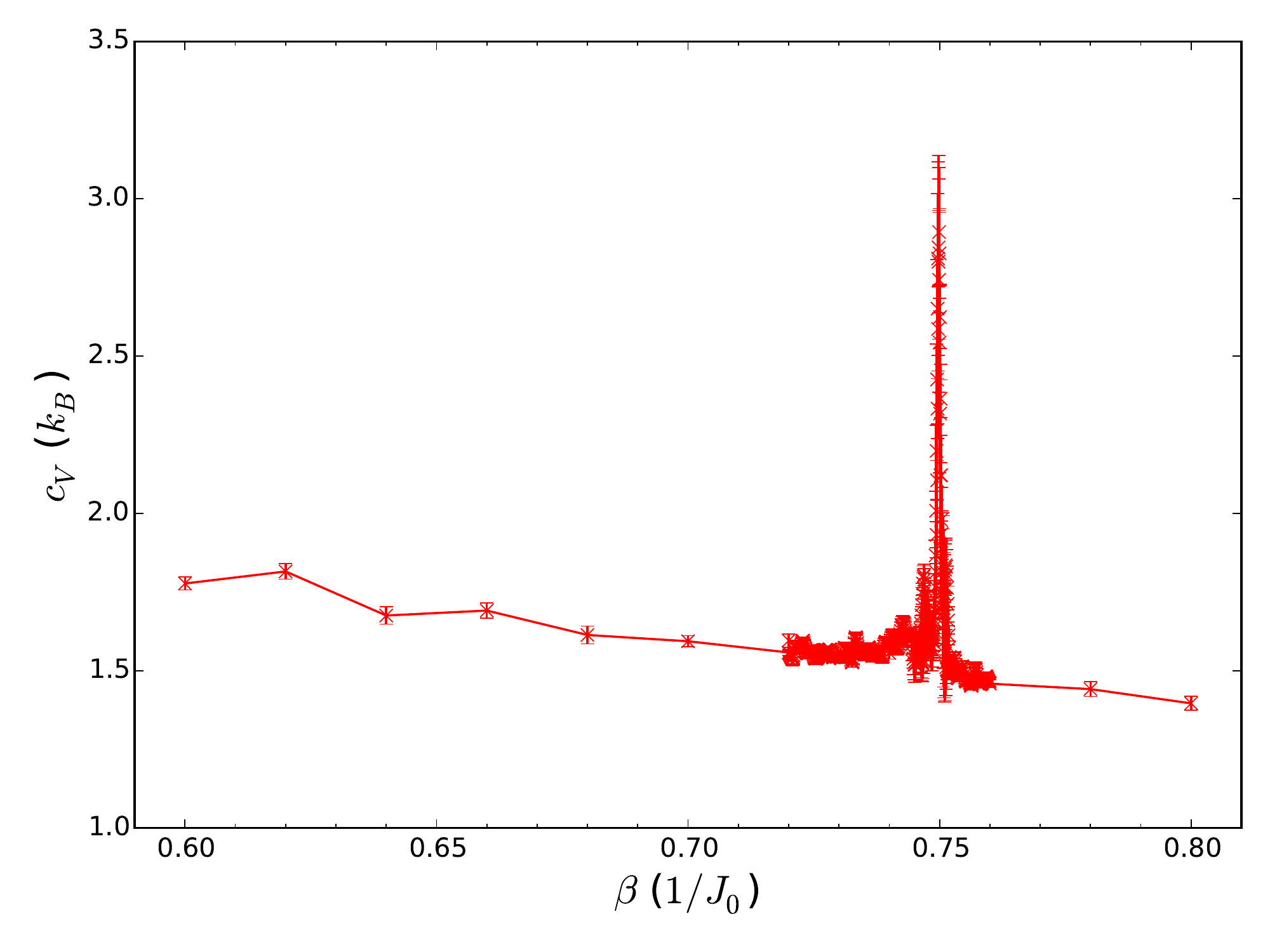}
\label{fig:1compheat}}\\
\subfloat{
\includegraphics[width=\columnwidth]{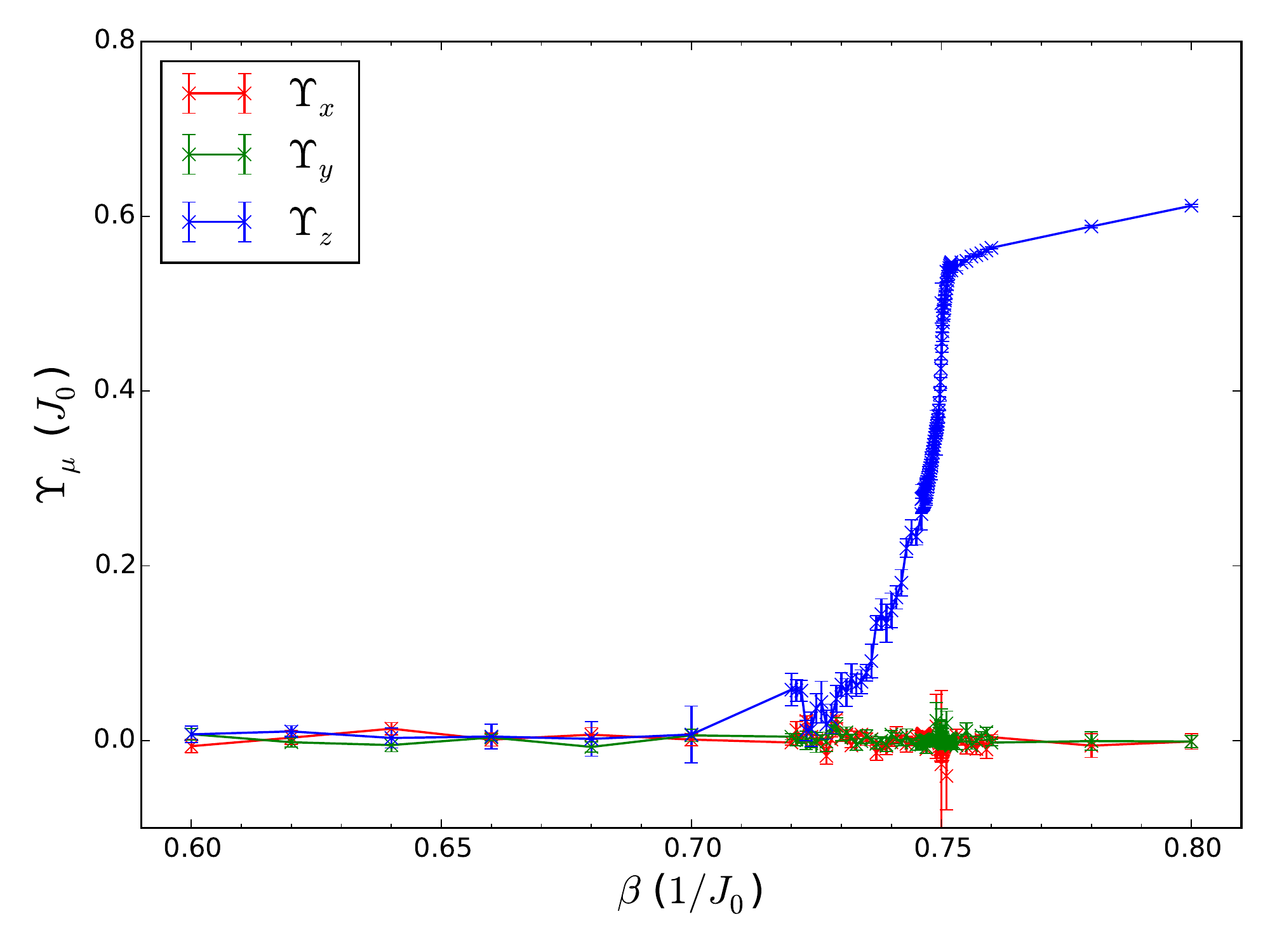}
\label{fig:1compheli}}
\caption{(Color online) Specific heat (top) and helicity moduli (bottom) for $N=1$, $f=1/16$, and
$\eta=10$. At $\beta=0.751$ we see a clear anomaly in the specific heat accompanied by a sharp jump
in the longitudinal helicity modulus. The transverse moduli remain zero throughout the transition.}
\label{fig:1comptrans}
\end{figure}

\begin{figure}
\centering
\includegraphics[width=\columnwidth]{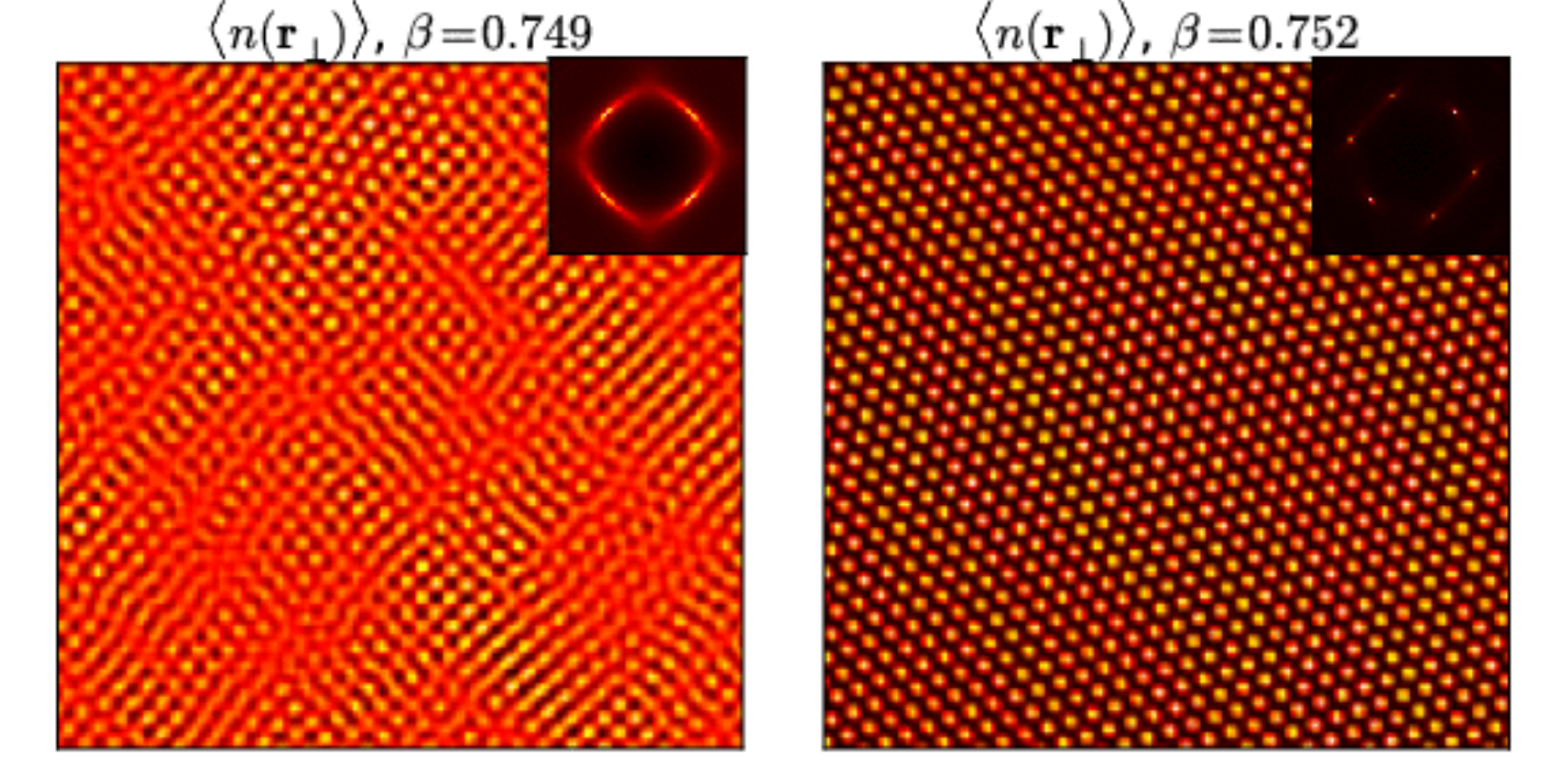}
\caption{(Color online) Vortex density, $n(\rvec_\perp)$, and structure factor, $S(\mathbf{q}_\perp)$, (inset) for
  $N=1$, $f=1/16$, and $\eta=10$, at inverse temperatures
$\beta=0.749$ (left) and $\beta=0.752$ (right). This corresponds to temperatures slightly higher and
lower, respectively, than the transition point, $\beta=0.751$.}
\label{fig:1compvortsf}
\end{figure}

The simulations were performed with $\eta=10$.  Fig.~\subref*{fig:1compheat} shows the specific
heat, which has strong signs of an anomaly at $\beta=0.751$. Fig.~\subref*{fig:1compheli} shows
that the anomaly in the specific heat is accompanied by a relatively sharp jump in the helicity
modulus in the $z$-direction. It is also important to note that the helicity moduli in the
transverse directions remain zero throughout the transition. This indicates that the vortex lattice
melts in a genuine phase transition, and not as a result of thermal depinning from the underlying
numerical lattice. This is therefore a strong indication of a first-order melting transition.
Fig.~\ref{fig:1compvortsf} shows the vortex density and structure factor immediately before and
after the transition. The high-temperature side shows an incoherent vortex liquid, characterized by
a circular structure factor. The-low temperature side shows that a clear hexagonal structure is
established as soon as the liquid freezes.

We emphasize that these results are not unexpected. The purpose of including them here, is to demonstrate that the method
of including amplitude fluctuations into the computation of the vortex lattice melting reproduces the known result for $N=1$,
previously obtained in the absence of amplitude fluctuations\cite{Teitel91,Teitel97,Hu97,Nguyen98_1,Nguyen98_2,Stroud98,Nguyen99,Chin99,Nguyen_EPL99},
but generally believed to be correct also when amplitude fluctuations are included.

\section{First order lattice melting for $N=2$} \label{sec:latticemelting}
{We next consider the melting transition for $N=2$, where the intercomponent density-density interaction
$2(\eta-\omega)\left|\psi_1\right|^2\left|\psi_2\right|^2$-term in the potential energy in  Eq. \ref{eq:inter_intra_g}  comes into play.
We consider the $U(1) \times U(1)$-symmetric case, i.e. $\omega \neq 0$. Again, the full spectrum of amplitude fluctuations is
included, using the methods described in Section~\ref{sec:MCdetails}.}

\begin{figure}
\centering
\subfloat{
\includegraphics[width=\columnwidth]{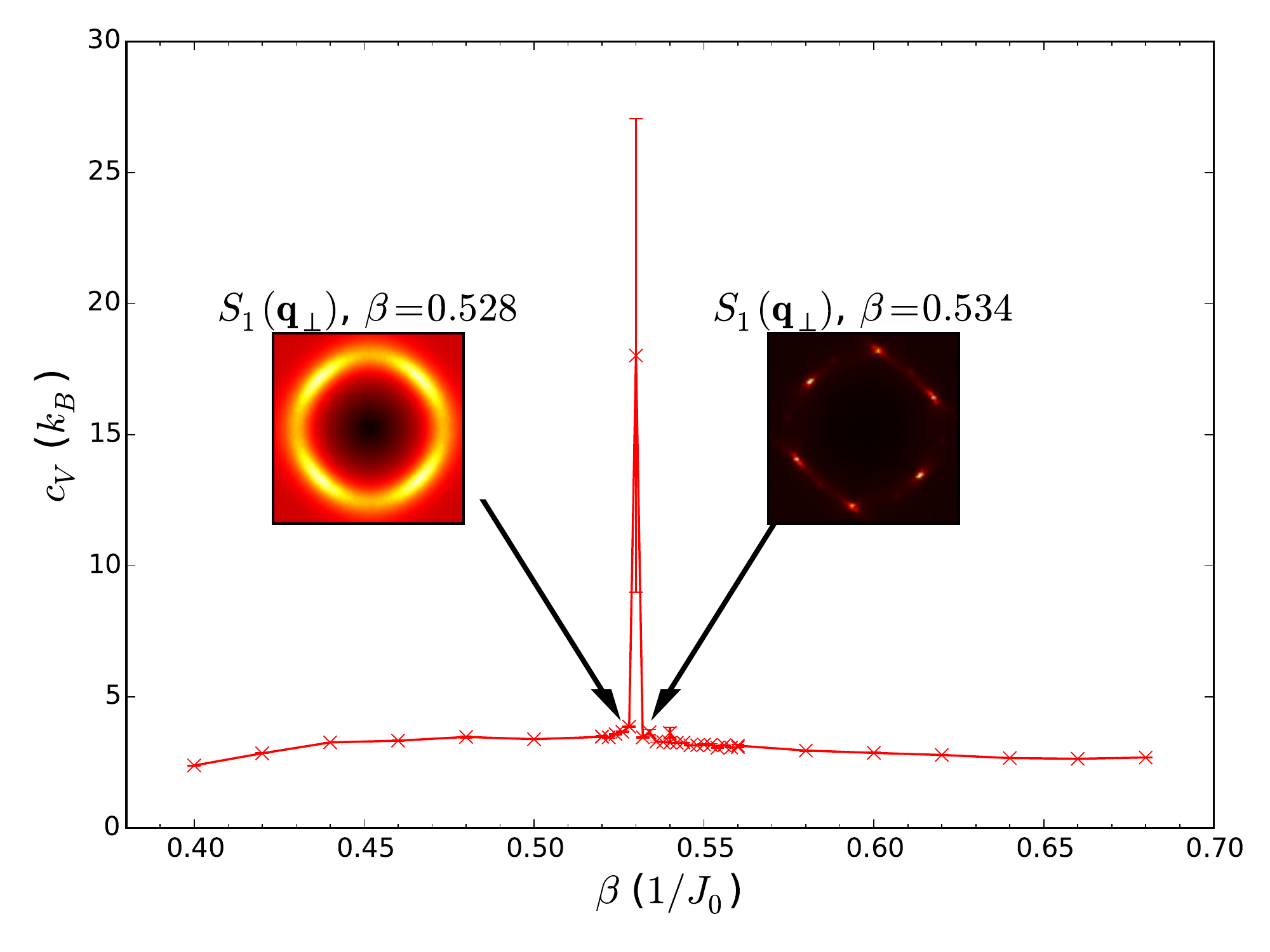}
\label{fig:2compheat05}}\\
\subfloat{
\includegraphics[width=\columnwidth]{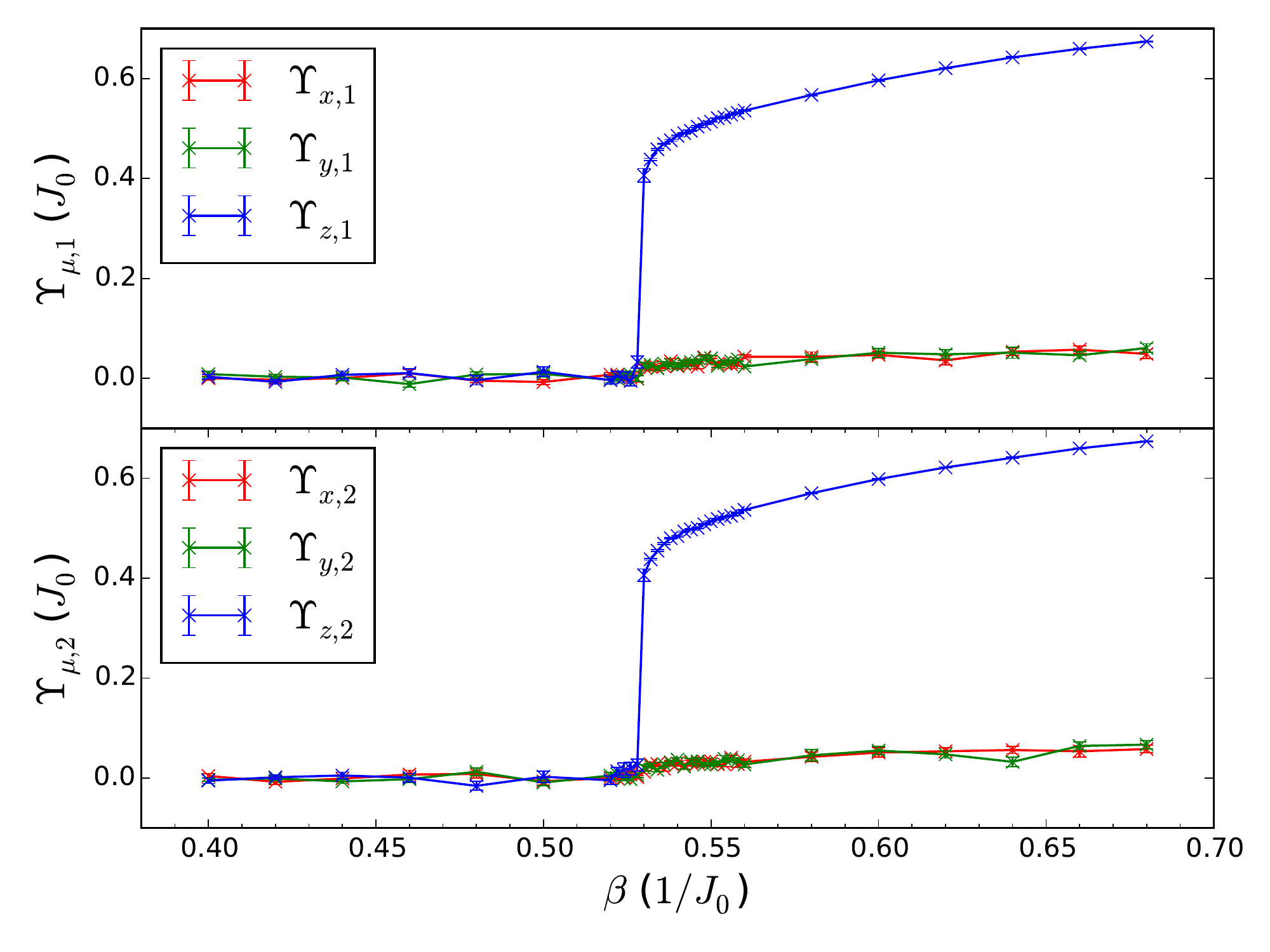}
\label{fig:2compheli05}}
\caption{(Color online) Specific heat (top) and helicity moduli of both components (bottom), for $N=2$,
    $f=1/16$, $\omega=1.0$ and $\eta=0.5$. At $\beta\approx 0.53$ we see a clear anomaly in the
    specific heat accompanied by a sharp jump in the longitudinal helicity moduli of both
    components. The transverse helicity moduli remain at zero throughout the transition. The insets
    in (a) show the structure factors at the high- and low-temperature sides of the transition,
    respectively $\beta=0.528$ and $\beta=0.534$. This clearly shows that the sharp anomaly in the
  specific heat separates an isotropic phase from a phase with hexagonal order.}
\label{fig:2comptrans05}
\end{figure}

For parameters $(\eta,\omega,f)=(0.5,1.0,1/16)$ and $(\eta,\omega,f)=(2.0,1.0,1/16)$ the lattices are clearly hexagonal and square,
respectively. The hexagonal lattice obtained for $\eta=0.5$ and $\omega=1.0$ was found to have a melting transition at $\beta\approx0.53$.
Fig.~\subref*{fig:2compheat05} shows the specific heat with a delta-function like anomaly at this temperature. Around this point,
we have used a closely spaced set of temperatures, in order to get a proper resolution of the anomaly.  Fig.~\subref*{fig:2compheli05}
shows the helicity moduli of both components. Both of the $z$-directed stiffnesses have a zero expectation value in the disordered
phase, indicating no phase coherence. In the ordered phase, both of $\langle\Upsilon_{z,i}\rangle$ develop finite expectation values which
means that the system has superfluidic properties along the direction of rotation. The two phases are divided by a sharp jump
in the longitudinal phase stiffness, a characteristic of a first-order transition. The drop is even sharper than what was obtained
for the $N=1$ case, indicating an even larger latent heat associated with the transition. The $x$- and $y$- directed stiffnesses remain
zero in the ordered state, which rules out any possibility of numerical pinning effects.\cite{Teitel94,Wheatley94}
Looking further at the insets of Fig.~\ref{fig:2compheat05}, which show the structure factors in the
disordered and ordered phase, we see clear evidence of an incoherent vortex liquid at $\beta<0.53$
in the left inset, while the right inset shows an ordered hexagonal vortex liquid lattice at $\beta>0.53$.

\begin{figure}
\centering
\subfloat{
\includegraphics[width=\columnwidth]{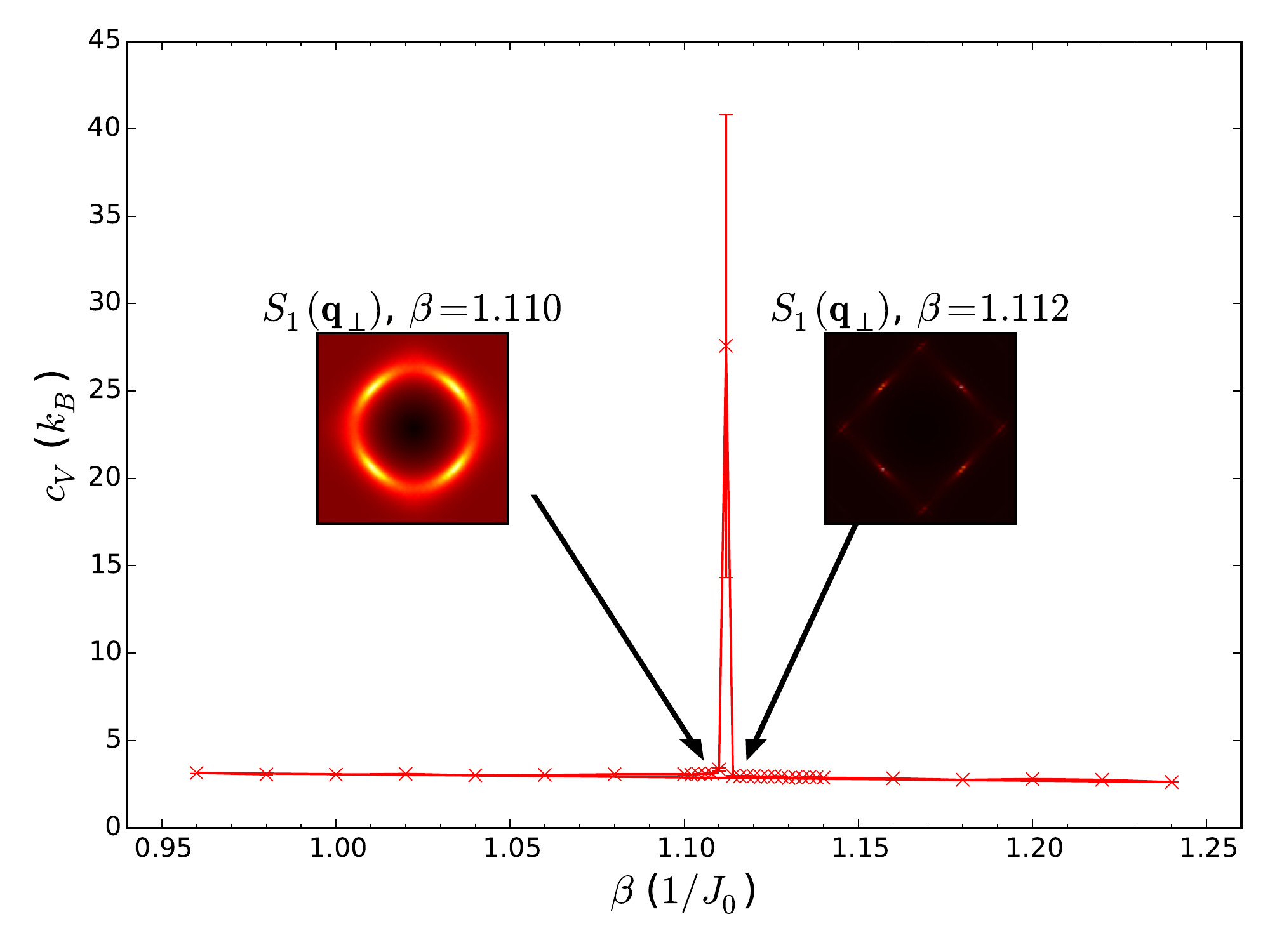}
\label{fig:2compheat2}}\\
\subfloat{
\includegraphics[width=\columnwidth]{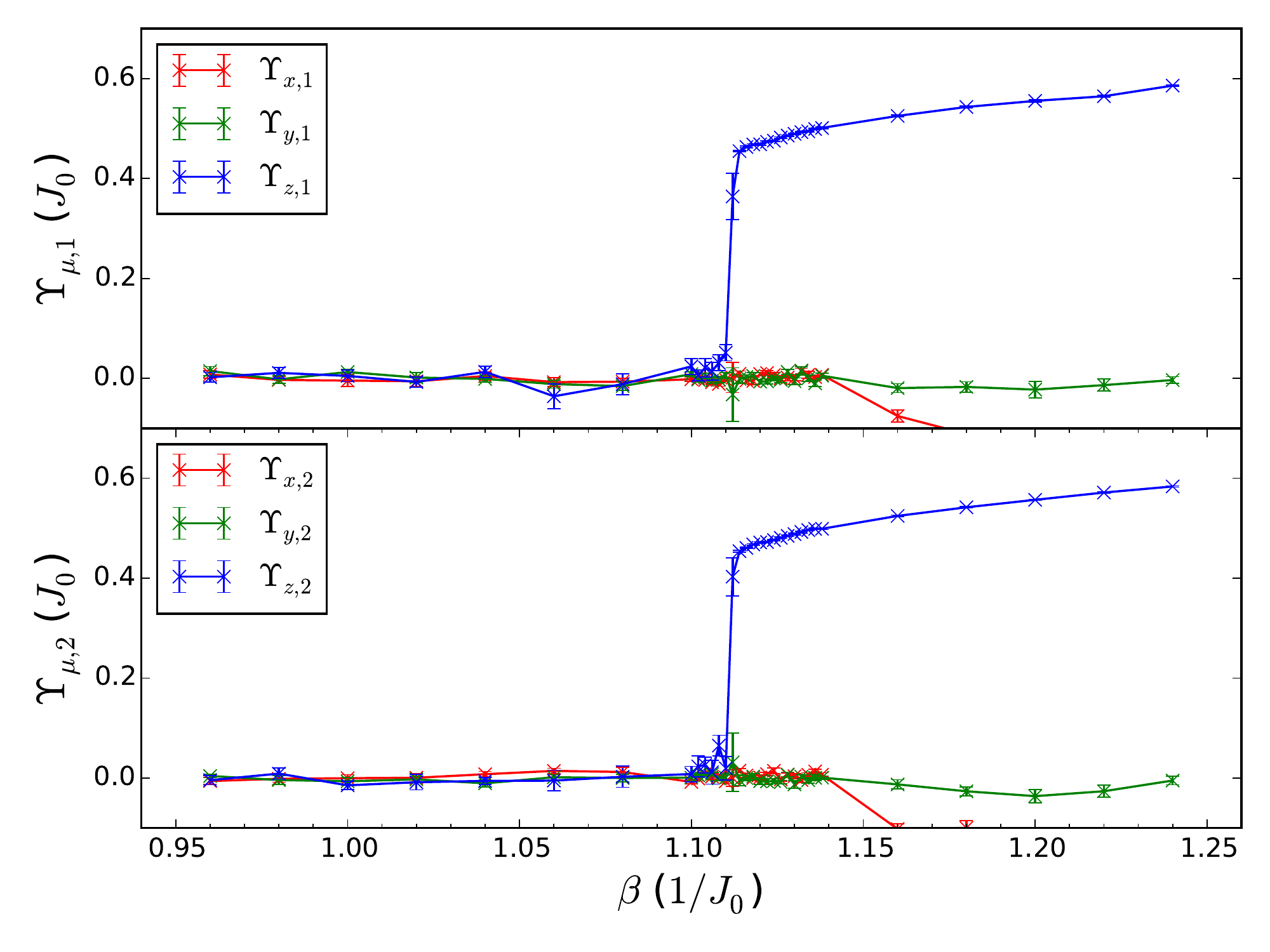}
\label{fig:2compheli2}}
\caption{(Color online) Specific heat (top) and helicity moduli of both components (bottom), for $N=2$,
    $f=1/16$, $\omega=1.0$ and $\eta=2.0$. At $\beta\approx 1.11$ we see a clear anomaly in the
    specific heat accompanied by a sharp jump in the longitudinal helicity moduli of both
    components. The transverse helicity moduli remain at zero throughout the transition, except for
    the $x$-directed modulus which drops to a negative value at a point well separated from the
    transition. The insets in (a) show the structure factors at the high- and low-temperature sides
  of the transition, respectively $\beta=1.110$ and $\beta=1.112$. This clearly shows that the sharp
anomaly in the specific heat separates an isotropic phase from a phase with square order.}
\label{fig:2comptrans2}
\end{figure}

Turning to the square lattice, now the parameters in question are $\eta=2.0$ and $\omega=1.0$. The
transition point is located at $\beta\approx1.11$. Fig.~\subref*{fig:2compheat2} shows the specific
heat. Again, an anomaly is located at the transition point. The helicity moduli, shown in
Fig.~\subref*{fig:2compheli2}, also show first order behavior. Both $z$-directed components are zero
on the high temperature side, and develop a finite value through a sharp jump at the low temperature
side. It is important to also consider the transverse components. Both
$\langle\Upsilon_{x,i}\rangle$ and $\langle\Upsilon_{y,i}\rangle$ are zero throughout the area
of interest. Here we note that the $x$-directed modulus drops to a tiny negative value at a point
after the transition. This is a non-physical effect, most likely caused by a metastable state. We
believe this is simply a numerical artifact, as we used a lower amount of Monte Carlo time away from
the transition. Turning our attention to the structure factors, shown in the insets of
Fig.~\ref{fig:2compheli2}, we again see the isotropic vortex liquid in the disordered side of the
transition, the ordered side shows a square four-fold symmetry.

Thus, both the square and the hexagonal lattices undergo first order melting transitions from their
respective ordered phases, into an isotropic vortex line liquid.

\section{Inter-component interaction, and its effect on density- and vortex-lattices.}
\label{app:tableaux}

In this appendix, we include more detailed figures of the vortex- and density-structures  in real-
and reciprocal space, as the inter-component interaction $2(\eta - \omega) |\psi_1|^2 |\psi_2|^2$ is
varied, to supplement the points made in Sections~\ref{sec:nwphases} and~\ref{sec:su2lattices}.

\begin{figure*}
    \centering
    \includegraphics[width=2\columnwidth]{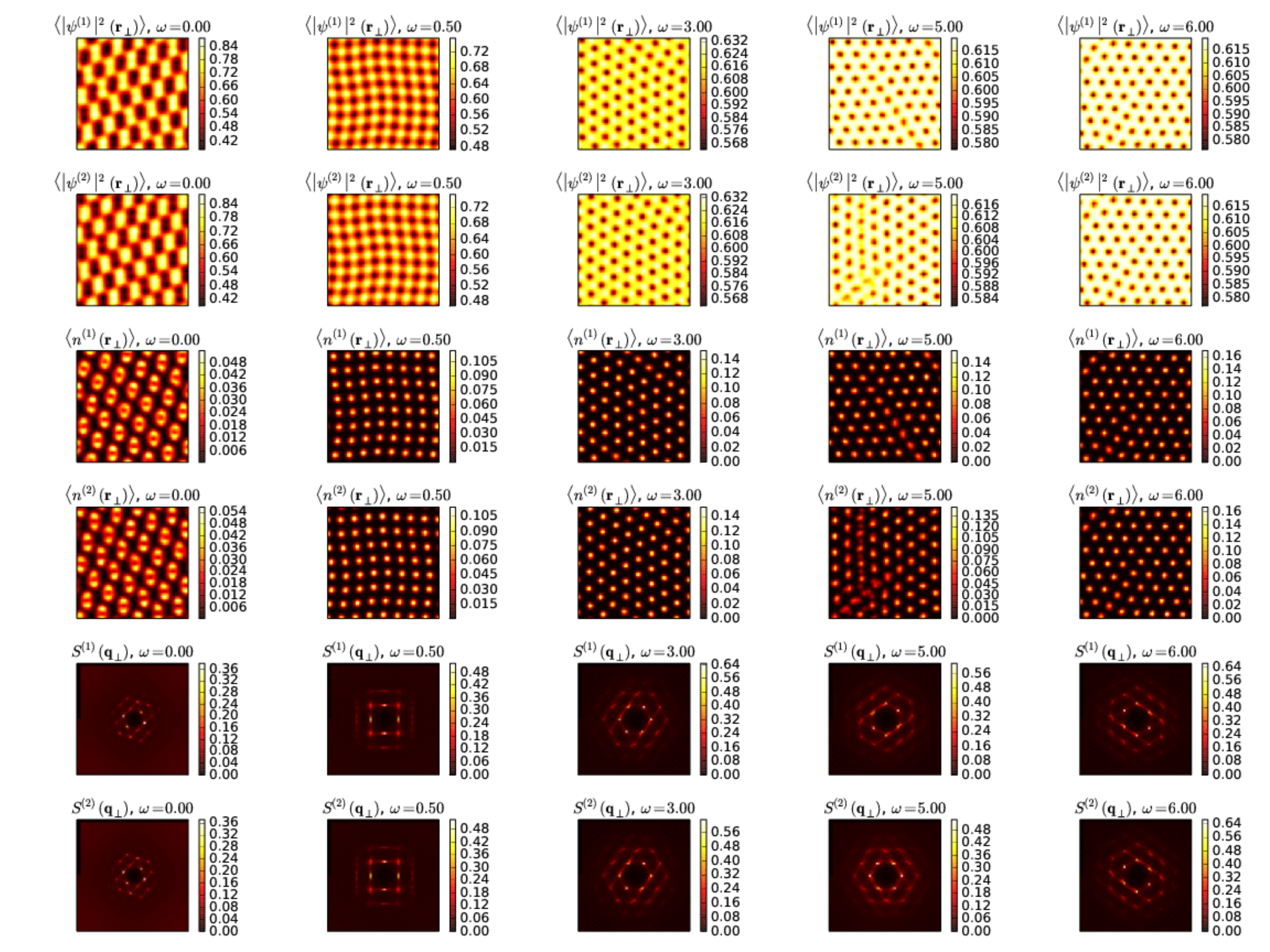}
    \caption{(Color online) Tableau illustrating the different density and vortex lattices in real space, as the
       parameter $\omega$ increases, \textit{i.e.}\, as the inter component density-density
       interaction $2(\eta-\omega) |\psi_1|^2 |\psi_2|^2$ decreases. This interaction promotes a
       square density- and vortex lattice. The parameters $f$, $\beta$, and $\eta$ are fixed to
       $f=1/64$, $\beta=1.5$ and $\eta=5$ while $\omega$ is increased from $0.0$ to $6.0$
       horizontally. The six rows show, from top to bottom the amplitude densities of components 1 and
       2, the vortex densities of components 1 and 2, and the structure factors of components 1 and 2.
       For $\omega=0$, which is the $\mathrm{SU(2)}$-symmetric case, the system exhibits a dimerized
       phase in component 1, which is complementary to a dimerized phase in component 2, shifted
       with respect to that of component 1 by an amount corresponding to the lattice constant of the
       density lattice. The ground state, where the roles of components 1 and 2 are interswitched, is
       degenerate with the illustrated phase.  Note that an area of the system with a high vortex
       density always corresponds to an area with a low amplitude density.  For the
       $\mathrm{SU(2)}$-symmetric case, $\mathrm{U(1)}$-vortices are not topological. When $\omega
       \neq 0$, the $\mathrm{SU(2)}$ symmetry is broken down to $\mathrm{U(1)} \times
       \mathrm{U(1)}$, and $\mathrm{U(1)}$ vortices are topological. The reduction of the
       interaction $2(\eta-\omega) |\psi_1|^2 |\psi_2|^2$ reduces the tendency towards formation of
       square density- and vortex-lattices, leading to an eventual reconstruction to a standard
       hexagonal vortex lattice, and hence a hexagonal density lattice. }
\label{fig:tableaux_l5}
\end{figure*}

\begin{figure*}
    \centering
    \includegraphics[width=2\columnwidth]{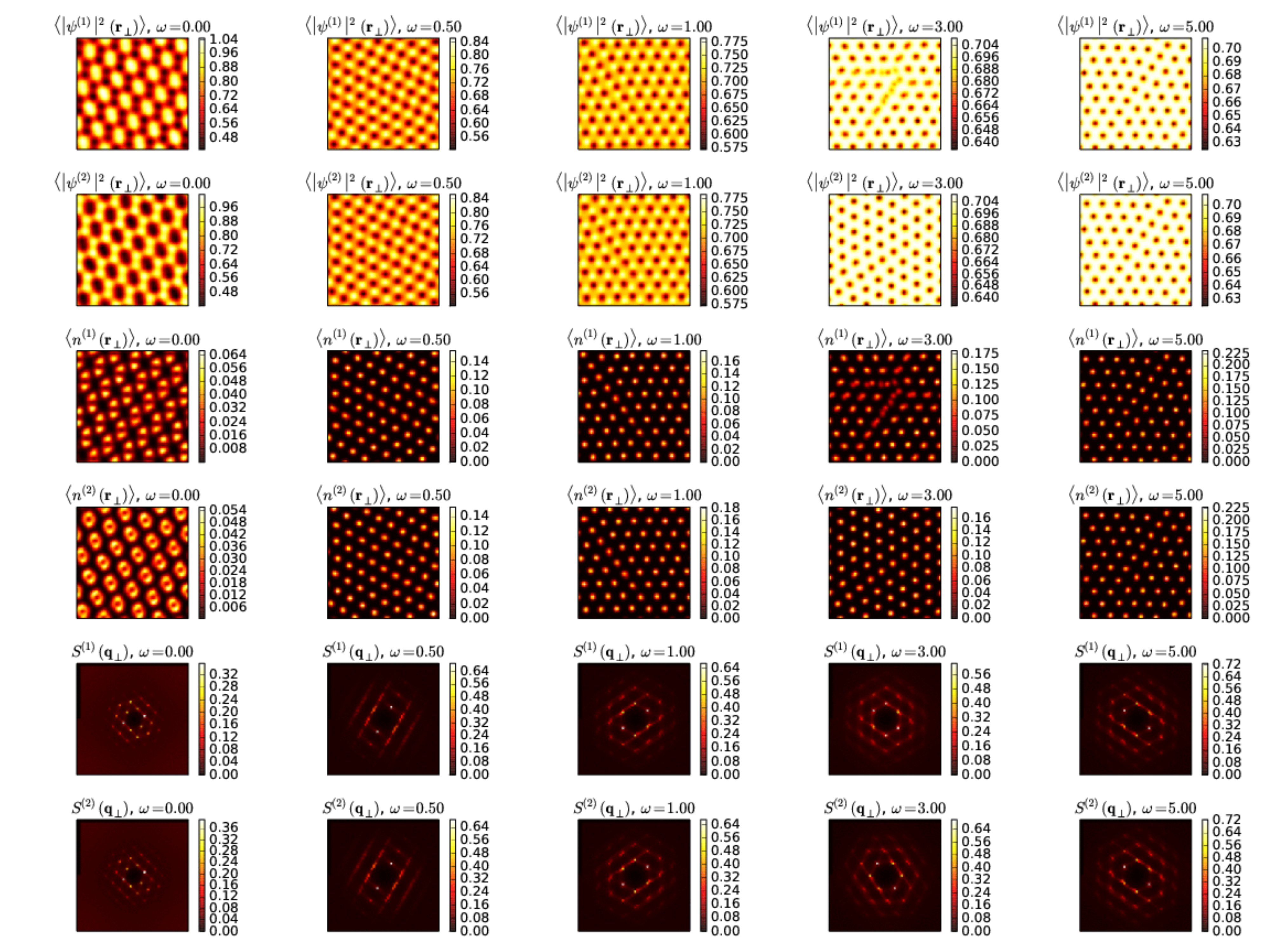}
    \caption{(Color online) Tableau illustrating the different density and vortex lattices in real space, as the
       parameter $\omega$ increases, \textit{i.e.}\ as the inter-component density-density
       interaction $2(\eta-\omega) |\psi_1|^2 |\psi_2|^2$ decreases. This interaction promotes a
       square density- and vortex lattice. The parameters $f$, $\beta$, and $\eta$ are fixed to
     $f=1/64$, $\beta=1.5$ and $\eta=3$ while $\omega$ is increased from $0.0$ to $5$
   horizontally. The six rows show, from top to bottom the amplitude densities of components 1 and 2,
the vortex densities of components 1 and 2, and the structure factors of components 1 and 2.}
\label{fig:tableaux_l3}
\end{figure*}

Figs.~\ref{fig:tableaux_l5} and~\ref{fig:tableaux_l3} illustrate how the vortex lattice, and the
{component densities } reconstruct as the inter-component density-density interaction $(\eta -
\omega) |\psi_1|^2 |\psi_2|^2$ changes. We do this by fixing $\eta$ at $5.0$ and $3.0$, respectively,
and tuning $\omega$. The inverse temperature, $\beta$, is also fixed in both tableaux. Common in
both figures is that the vortices first form two interlaced square lattices for sufficiently small
$\omega$, and, by extension, large inter-component coupling. Then the lattices reconstruct into a
hexagonal structure. Note that the hexagonal lattices of the two components start out slightly
shifted with respect to each other, but becomes completely cocentered when $\omega > \eta$. This
final state corresponds to an attractive inter-component coupling.

The behavior of the amplitude densities is explained in Sec.~\ref{sec:nwphases}, and we can
compare the reasoning to the top two rows of Figs.~\ref{fig:tableaux_l5}
and~\ref{fig:tableaux_l3}. First of all, the presence of a vortex locally suppresses the amplitude,
which again may affect the immediate neighborhood, depending on the value of the inter-component
coupling. For strong repulsive couplings a suppression of the amplitude of one color in an area
leads to an enhancement of the amplitude of the other color in the same area. The absence of
vortices in the neighborhood then leads to the opposite effect. This causes staggering of the
amplitude densities and formation of  distinct vortex sublattices. Considering carefully the range of
variation in the amplitudes, it is seen that there are rather large gradients for the square
structures. When the coupling is only weakly repulsive or even attractive there is a much less
dramatic effect. The variations in the amplitudes are much smaller; there is little to no
staggering.

\begin{figure*}
    \centering
    \includegraphics[width=2\columnwidth]{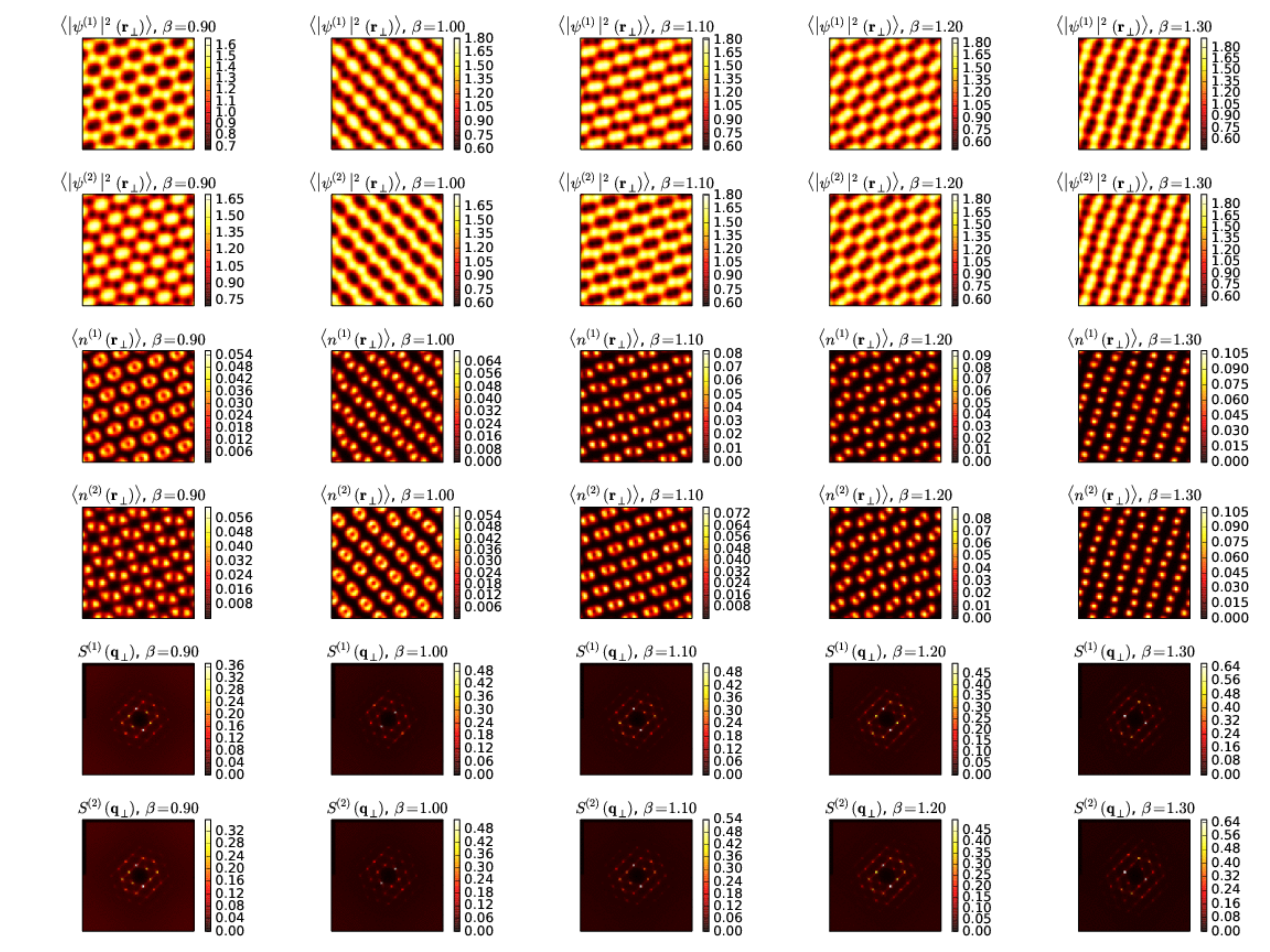}
    \caption{(Color online) Tableau illustrating the different $\mathrm{SU(2)}$ density and vortex lattices in
      real space, as the parameter $\beta$ increases. The parameters $f$, $\eta$, and $\omega$ are
      fixed to $f=1/64$, $\eta=1.0$ and $\omega=0.0$ while $\beta$ is increased from $0.9$ to $1.3$
      horizontally. The six rows show, from top to bottom the amplitude densities of components 1 and
      2, the vortex densities of components 1 and 2, and the structure factors of components 1 and 2.
    Note how the vortex structures and the density structures always track, that is, an area of the
  system with a high vortex density always corresponds to an area with a low amplitude density.}
\label{fig:tableaux_su2}
\end{figure*}

The first column of Figs.~\ref{fig:tableaux_l5} and~\ref{fig:tableaux_l3} is in a different class from
the rest. Here $\omega=0$, and we are in the $\mathrm{SU(2)}$ regime.  Fig.~\ref{fig:tableaux_su2}
further illustrates the wide variety of ground states obtainable here.  This tableau, in contrast to
the two previous ones, has a fixed $\eta$ and $\omega$, while we vary the inverse temperature $\beta$
from column to column. These pictures are all taken from a single simulation, evolved through Monte
Carlo sampling from a single randomized initial state as $\beta$ is increased. The vortex lattice
initially forms at around $\beta=0.7$, and evolves continuously.
It continues to evolve even at the lowest temperatures ($\beta=1.5$) used in the simulation. This
pattern is common in all simulations done with similar parameter sets.

The common features in the $\mathrm{SU(2)}$ lattices are clearly seen in
Fig.~\ref{fig:tableaux_su2}. The vortices tend to form dimers, which usually have some global
alignment. The alignment is evident in the Bragg peaks, as we in most cases have two opposing peaks
of higher intensity than the rest. The vortex dimer complexes always arrange themselves in a
hexagonal structure, which is also seen in the structure factors.

\bibliography{references}

\end{document}